\documentclass[11pt]{article}

\usepackage[preprint]{acl}

\usepackage{times}
\usepackage{latexsym}

\usepackage[T1]{fontenc}

\usepackage[utf8]{inputenc}

\usepackage{tcolorbox}

\usepackage{microtype}

\usepackage{inconsolata}

\usepackage{graphicx}

\usepackage{booktabs}
\usepackage{colortbl}
\usepackage[table]{xcolor}
\usepackage{array}
\usepackage{makecell}
\usepackage{graphicx}
\usepackage{float}
\usepackage{amsmath}
\usepackage{multirow}
\usepackage{amssymb}

\title{When Should Queries Be Decomposed? A Stage-Aware Study of Query Decomposition for Multi-Condition Retrieval}

\author{%
 \textbf{Bochao Yin\textsuperscript{1}\thanks{Equal contribution}},\;
 \textbf{Xuan Lu\textsuperscript{1,2}\protect\footnotemark[\value{footnote}]},\;
\textbf{Zhengyu Qi\textsuperscript{1}},\;
 \textbf{Xiaoyu Shen\textsuperscript{1,3}\thanks{Corresponding author}}\\[4pt]
 \textsuperscript{1}Eastern Institute of Technology, Ningbo
  \textsuperscript{2}Shanghai Jiao Tong University\\
 \textsuperscript{3}Zhejiang Key Laboratory of Industrial Intelligence and Digital Twin\\
 \texttt{lux1997@sjtu.edu.cn\quad xyshen@eitech.edu.cn}
}

\begin{document}
\maketitle
\begin{abstract}

Multi-condition retrieval requires systems to identify documents that satisfy multiple distinct constraints, moving beyond mere topical relevance. While query decomposition is widely adopted as an intuitive remedy, its effectiveness across different retrieval pipeline stages remains underexplored. In this paper, we conduct a stage-aware empirical study and uncover a stark, stage-dependent effect: decomposition during initial retrieval frequently harms retrieval performance due to semantic dilution, yet substantially improves reranking by enabling more fine-grained constraint verification. Motivated by these insights, we propose a principled \textbf{Stage-Aware Decomposition} framework that retains the monolithic query during initial retrieval to preserve global semantic context, while employing sub-queries exclusively during reranking for fine-grained constraint matching. Extensive evaluations on the MultiConIR and SSRB benchmarks demonstrate that our framework consistently improves ranking performance for compositional queries across multiple retrieval and reranking models. We release our code at \url{https://github.com/EIT-NLP/Query-Decompose}.

\end{abstract}

\section{Introduction}
Information Retrieval (IR) underpins modern information access by enabling systems to identify and rank relevant documents. Retrieval has evolved from lexical matching and structured filtering~\cite{DBLP:journals/ftir/RobertsonZ09,DBLP:conf/chi/YeeSLH03} to neural semantic retrieval~\cite{DBLP:conf/emnlp/KarpukhinOMLWEC20,DBLP:conf/sigir/KhattabZ20} and, more recently, LLM-based query reformulation and reranking~\cite{DBLP:conf/emnlp/0001YMWRCYR23,DBLP:conf/acl/GaoMLC23}. Despite these advances, many real-world queries remain challenging because they contain multiple conditions that must be considered jointly. For example, a query such as ``\textit{a lightweight laptop with 16GB RAM under \$1500 for gaming, and at least 512GB SSD storage}'' specifies several independent requirements. We refer to this setting as \emph{multi-condition retrieval}, where the goal is to rank documents according to how well they satisfy the query's multiple conditions collectively.


Recent benchmarks such as MultiConIR~\cite{lu-etal-2025-multiconir} and SSRB~\cite{zhang2025ssrb} show that modern retrievers struggle with such queries, particularly in distinguishing highly relevant documents from hard negatives that match the general topic but satisfy fewer specific conditions. A natural approach is \emph{query decomposition}, which breaks a complex query into simpler sub-queries corresponding to its constituent conditions. This idea has been explored in complex question answering~\cite{DBLP:conf/acl/MinZZH19}, retrieval-augmented frameworks~\cite{ammann-etal-2025-question}, and agentic search systems~\cite{DBLP:conf/emnlp/PressZMSSL23,DBLP:journals/corr/abs-2503-09516,DBLP:journals/corr/abs-2508-09303}.

However, despite its growing adoption, the effectiveness of query decomposition for multi-condition retrieval remains poorly understood. In particular, it is unclear whether decomposition is equally useful at different stages of a multi-stage retrieval pipeline, where initial retrieval aims to preserve recall while reranking focuses on fine-grained discrimination. This leads to a critical but underexplored research question: \textit{Does the effectiveness of query decomposition depend on where it is applied within the multi-stage retrieval pipeline?}
In this work, we conduct a systematic empirical study of query decomposition and uncover a stark, stage-dependent effect: \emph{decomposition frequently impairs initial retrieval, but improves reranking on the same queries}. Quantitatively, on complex queries, decomposition decreases retrieval NDCG@10 by 1.6--4.3 points across different retrievers, while improving reranking NDCG@10 by 3.9--7.6 points over standard reranking across different reranker models. We trace this asymmetry to a phenomenon we term \emph{semantic dilution}. A complex query combines multiple conditions that collectively provide a strong signal for identifying the most relevant documents. Decomposition strips away this conjunctive structure, leaving each sub-query with only a fragment of the original intent. Each sub-query therefore matches a much larger candidate pool, including documents that satisfy one condition but fail to satisfy the remaining ones. During initial retrieval, where the system must search a large corpus and preserve recall, this loss of global context weakens the discriminative signal of the original query. Partial matches can consequently be promoted over documents that satisfy substantially more of the original conditions.

In contrast, decomposition becomes beneficial during reranking. Rerankers operate over a small, pre-filtered candidate pool and can model fine-grained query--document interactions at the token level~\cite{lu2026comprankefficientllmreranking,fan2026minirerankerefficientmultimodalreranking}. At this stage, the goal is not broad candidate discovery but precise discrimination among already plausible candidates. Decomposed sub-queries make individual conditions explicit, allowing the reranker to assess how well each candidate matches different parts of the original intent. Thus, the same decomposition that weakens global retrieval signals can provide useful condition-level information when applied to reranking.


Motivated by these insights, we propose a principled \textbf{Stage-Aware Decomposition} framework that aligns query representations with the objectives of different retrieval stages. We retain the original monolithic query for initial retrieval, preserving the complete query intent and safeguarding global recall. We then apply query decomposition exclusively during reranking, where fine-grained interactions between candidate documents and individual sub-queries can be exploited for more precise ranking. This design combines the complementary strengths of the two representations: the monolithic query preserves global relevance during candidate generation, while decomposed sub-queries expose individual conditions for fine-grained reranking.

To summarize, our primary contributions are:
\begin{itemize}
\item \textbf{Stage-Dependent Insights}: We systematically evaluate query decomposition across different stages of multi-condition retrieval and identify \emph{semantic dilution} as a key cause of performance degradation during the initial retrieval stage.
\item \textbf{Stage-Aware Framework}: We introduce a framework that strategically aligns query with pipeline stages, preserving the monolithic query for initial retrieval while exploiting decomposed sub-queries for reranking.
\item \textbf{Extensive Evaluation}: Through comprehensive evaluations on the MultiConIR and SSRB benchmarks across various embedding and reranking models, we demonstrate that our approach substantially and consistently enhances ranking precision for complex multi-condition queries.
\end{itemize}

\section{Related Work}
\paragraph{Multi-condition retrieval}
Traditional retrieval benchmarks such as MS MARCO~\cite{DBLP:conf/nips/NguyenRSGTMD16} and BEIR~\cite{DBLP:journals/corr/abs-2104-08663} have driven the development of modern retrieval models, but they mainly evaluate retrieval for individual natural language queries based on lexical or semantic similarity. 
Such settings often fail to capture user intent involving multiple independent constraints.
Recent benchmarks like BRIGHT~\cite{DBLP:conf/iclr/SuYXSMWLSST0YA025} introduce reasoning-intensive queries beyond surface matching. 
Nevertheless, retrieval is still framed as satisfying a single unified query objective. 
More recent work explicitly studies multi-condition retrieval. BIRCO~\cite{DBLP:journals/corr/abs-2402-14151} targets retrieval tasks with multi-faceted objectives, MultiConIR~\cite{lu-etal-2025-multiconir} evaluates retrieval robustness as query conditions increase, MCMR~\cite{lu2026beyond} evaluates multi-condition multimodal retrieval, focusing on whether retrieval models can satisfy multiple fine-grained semantic constraints across modalities, and 
SSRB~\cite{zhang2025ssrb} constructs a large-scale benchmark for natural language queries that combine exact matching constraints and fuzzy semantic requirements across multiple fields.
These studies highlight the increasing difficulty of satisfying multiple constraints simultaneously, indicating that existing retrieval models still struggle with complex query requirements~\cite{huang2026mmeb, han2026makesgoodinstructiontuningdata}.
This gap motivates further research on robust multi-condition retrieval. In particular, it calls for more effective ways to handle complex queries with multiple constraints, which we discuss in the following subsection.

\paragraph{Query Decomposition}

Query decomposition has been widely studied as a strategy for handling complex information needs. 
Prior work on query reformulation and rewriting has been shown to improve retrieval effectiveness,  particularly in conversational search and information retrieval settings~\cite{DBLP:journals/tois/LinYNTWL21,DBLP:conf/emnlp/NogueiraC17}. 
Building on this idea, query decomposition splits a complex query into simpler sub-queries, and has been applied in multi-hop reasoning and retrieval-augmented generation (RAG), where decomposed queries can improve evidence retrieval and answer accuracy~\cite{DBLP:journals/tacl/TrivediBKS22,lin-etal-2023-decomposing,ammann-etal-2025-question,fan2025visipruner,fan2026visual}.
More recent work further optimizes decomposition to better align with downstream retrievers or dynamically select or prioritize sub-queries during retrieval~\cite{DBLP:conf/icml/LiuLWC25,DBLP:journals/corr/abs-2510-18633}. 
Decomposition has also been explored in structured retrieval settings, where sub-queries are used to capture fine-grained attributes or sub-intents~\cite{zhu-etal-2025-hint,DBLP:journals/corr/abs-2509-06544}.
Despite these advances, existing approaches often treat decomposed sub-queries independently and focus on improving recall. 
This limits their ability to model global consistency across multiple constraints, particularly in multi-condition retrieval.
In contrast, we study query decomposition from a stage-aware perspective and show that its effectiveness critically depends on how it is integrated into retrieval and reranking pipelines.

\section{Problem \& Experimental Setup}

\subsection{Task Definition}



\paragraph{Multi-Condition Retrieval}
We study information retrieval under multi-condition queries, where a query specifies multiple constraints that must be jointly satisfied. Formally, we represent a query as a set of atomic conditions $Q = \{c_1, c_2, \dots, c_k\}$. Given a document $d$, we say $d$ is \textit{fully relevant} to $Q$ if it satisfies all $k$ conditions simultaneously; otherwise it is partially relevant or irrelevant. The retrieval objective is to rank fully relevant documents above all others.

\paragraph{Query Decomposition}
To model compositional queries, we consider query decomposition, which transforms a query $Q$ into a set of sub-queries via a decomposition function $\mathcal{D}$:
\begin{equation}
\mathcal{D}(Q) = \{q_1, q_2, \dots, q_m\}.
\end{equation}
Each sub-query $q_i$ is associated with a set of conditions $C(q_i) \subseteq \{c_1, \dots, c_j\}$. Collectively, these sub-queries provide partial views of the original query, each capturing a subset of the constraints.

\paragraph{Two-Stage Retrieval Pipeline}
We consider a standard two-stage retrieval pipeline. In the first stage, a retriever produces a candidate set $\mathcal{C}_K$ of size $K$ from the corpus. In the second stage, a reranker assigns fine-grained relevance scores to each candidate in $\mathcal{C}_K$ and returns a reordered list. At each stage, the input query can be either the original query $Q$ or a set of decomposed sub-queries $\mathcal{D}(Q)$. Our central question is whether the effectiveness of query decomposition depends on the stage at which it is applied.

\paragraph{Dataset}
We evaluate on the MultiConIR benchmark~\cite{lu-etal-2025-multiconir}. which covers five domains—Books, Movies, People, Medical Cases, and 
Legal Documents—with queries containing $k \in \{1, \dots, 10\}$ conditions. 
Each query is paired with a positive document $d^+$ satisfying all $k$ 
conditions and graded hard negatives $\{d_j\}_{j=0}^{k-1}$, where $d_j$ 
satisfies exactly $j$ conditions.




\subsection{Evaluation Metrics}

We evaluate retrieval and ranking performance using \textbf{Recall@$K$}, \textbf{NDCG@$K$}, and \textbf{Win Rate}. 
Recall evaluates whether relevant documents are preserved during the initial retrieval stage;
NDCG is highly suitable for tasks involving graded relevance judgment~\cite{DBLP:journals/corr/abs-2104-08663}; 
Win Rate measures the model's fine-grained discrimination capability~\cite{lu-etal-2025-multiconir}.

The Win Rate is the probability that the true positive document $d^+$ scores higher than its hardest negative $d^-_{*}$, defined as the incorrect document that satisfies the largest number of query conditions. 
Formally, over a set of queries $Q$:

\begin{equation}
\text{Win Rate} = \frac{1}{|Q|} \sum_{q \in Q} \mathbf{1} \big( S(q, d^+) > S(q, d^-_{*}) \big)
\end{equation}

where $S(q, d)$ denotes the relevance score predicted by the model for a query-document pair.

Detailed Win Rate results are provided in Appendix~\ref{sec:appendix}.

\begin{figure*}[h]
    \centering
    \includegraphics[width=\textwidth]{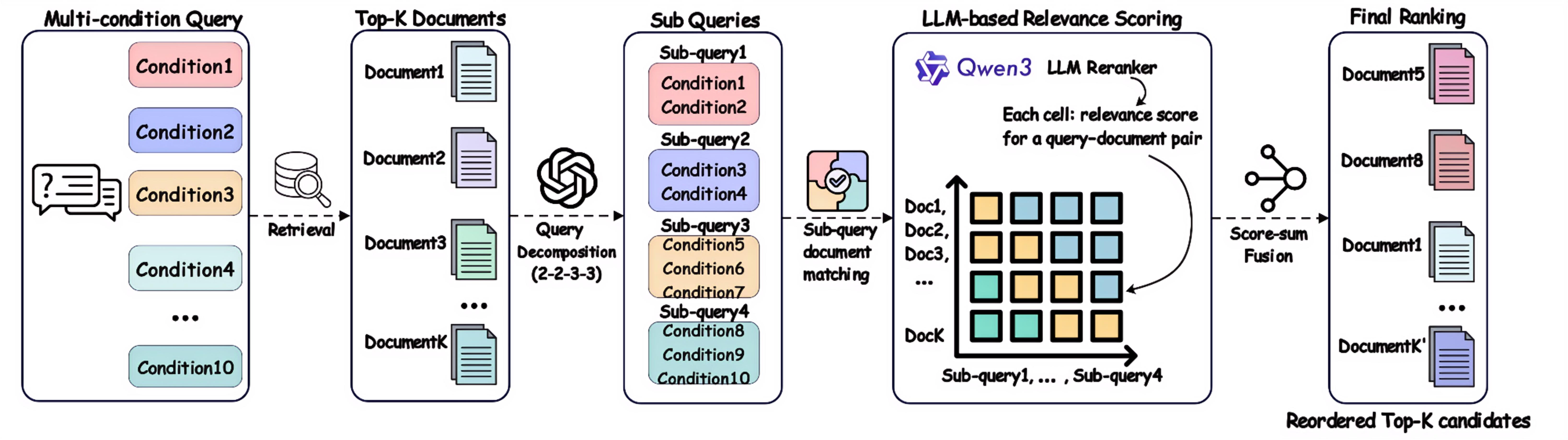}
    \caption{
    Overview of the Stage-Aware Decomposition Framework
    }
    \label{fig:rerank_pipeline}
    \vspace{-4mm}
\end{figure*}

\subsection{Retrieval and Reranking Setup}
\paragraph{Retrieval Setup}
We evaluate the effect of query decomposition on retrieval performance across five datasets. We use two representative embedding model architectures: (1) BGE-large-en-v1.5~\cite{DBLP:conf/sigir/XiaoLZMLN24}, an encoder-only model based on BERT~\cite{DBLP:conf/naacl/DevlinCLT19}; and (2) the Qwen3-Embedding series~\cite{DBLP:journals/corr/abs-2506-05176}, a decoder-only\cite{DBLP:conf/nips/VaswaniSPUJGKP17} model family built on LLMs.

For the Qwen3-Embedding series, we select variants of different sizes (0.6B, 4B, and 8B). For decomposed retrieval, we retrieve the top-1000 candidates for each sub-query and assign a score of zero to all documents beyond this cutoff before fusion. For this retrieval stage, we retrieve top-50 candidates then passed to the reranking stage.

\paragraph{Reranking Setup}
We adopt a pointwise reranking paradigm~\cite{lu2026rethinking,lu2025toolsunderdocumentedsimpledocument}, where each candidate document is independently evaluated against each decomposed sub-query.
For reranking models, we use bge-reranker-v2-m3~\cite{li2023making,chen2024bge} and the Qwen3-Reranker series~\cite{DBLP:journals/corr/abs-2506-05176} (0.6B, 4B, and 8B). We frame reranking as a binary relevance classification task and compute scores using the normalized logits of the ``Yes'' and ``No'' tokens. Discussions on listwise reranking and token-based scoring experiments are provided in Appendix~\ref{app:logit_reranking}.

\section{Methodology}
Intuitively, query decomposition offers a natural way to improve multi-condition retrieval: decomposing a long, information-dense query into shorter sub-queries allows the model to be more precised about each condition. By aggregating the top candidates retrieved for each sub-query through a fusion mechanism, the final result better reflects documents that simultaneously satisfy all conditions.
\subsection{Query decomposition}
Within our experimental framework, we investigate the efficacy of query decomposition by introducing three strategies, which are formulated as follows:
\paragraph{Naive Decomposition}
In this approach, a complex query $Q$ is strictly partitioned into single-condition sub-queries. The decomposition function $\mathcal{D}_{\text{naive}}$ is formulated as:
\[
    \mathcal{D}_{\text{naive}}(Q) = \{q_i\}_{i=1}^{n} \quad \text{s.t.} \quad |C(q_i)| = 1, \, \forall i,
\]
where $n$ is the total number of conditions in $Q$, and $C(q_i)$ denotes the set of semantic conditions contained within sub-query $q_i$.

\paragraph{Fixed-size Decomposition}
In contrast to the naive approach, which restricts each sub-query to a single condition, this strategy partitions a complex query $Q$ into sub-queries that each contain between $k$ and $j$ conditions, where $1 \le k \le j \le n$ and $n$ denotes the total number of conditions in $Q$. The decomposition function $\mathcal{D}_{\text{fixed}}$ is formulated as:
\[
    \mathcal{D}_{\text{fixed}}(Q) = \{q_i\}_{i=1}^{m} \quad \text{s.t.} \quad k \le |C(q_i)| \le j, \, \forall i,
\]
where $m$ is the total number of resulting sub-queries and $C(q_i)$ denotes the set of semantic conditions contained within sub-query $q_i$.

\paragraph{Adaptive Decomposition}
Unlike the previous rule-based strategies, this approach leverages an LLM to dynamically determine the optimal granularity of decomposition. The decomposition function $\mathcal{D}_{\text{adaptive}}$, parameterized by the LLM (denoted as $\mathcal{M}$), is formulated as:
\[
    \mathcal{D}_{\text{adaptive}}(Q; \mathcal{M}) = \{q_i\}_{i=1}^{k} \quad \text{s.t.} \quad |C(q_i)| = d_i,
\]
where $d_i \ge 1$ denotes the number of semantic conditions dynamically assigned to the $i$-th sub-query, and $k$ is the total number of generated sub-queries. Both $d_i$ and $k$ are inferred by $\mathcal{M}$ based on the semantic context of the original query $Q$.

\subsection{Fusion Strategy}
After query decomposition, we perform independent retrieval/reranking for each sub-query. To aggregate the result, we perform two different fusion strategies: Reciprocal Rank Fusion (RRF)~\cite{DBLP:conf/sigir/CormackCB09} and Score-Sum Fusion.

\paragraph{Reciprocal Rank Fusion (RRF)}
For a given document $d$, its RRF score is computed by summing the inverse of its rank across all decomposed sub-queries $q_i \in Q_{\text{sub}}$:

\begin{equation}
\mathrm{RRF}(d) = \sum_{q_i \in Q_{\text{sub}}} \frac{1}{k + \mathrm{rank}_i(d)}
\end{equation}

where $\mathrm{rank}_i(d)$ denotes the rank position of document $d$ retrieved by sub-query $q_i$, and $k$ is a smoothing constant. The final candidate list is generated by sorting the documents in descending order of their $\mathrm{RRF}(d)$ scores. RRF fusion provides a highly robust, training-free mechanism by relying strictly on the rank permutations. We provide a sensitivity analysis of $k$ in Appendix~\ref{app:rrf_sensitivity}.

\paragraph{Score-Sum Fusion}
Rather than utilizing rank positions, this method computes the final relevance score of document $d$ by directly aggregating its raw retrieval or reranking scores $s_i(d)$ across all sub-queries:

\begin{equation}
\mathrm{Score}(d) = \sum_{q_i \in Q_{\text{sub}}} s_i(d)
\end{equation}

Similar to RRF, documents are subsequently sorted in descending order based on their aggregated $\mathrm{Score}(d)$ to produce the final ranked list. Score-Sum Fusion retains the absolute magnitude of relevance scores, ensuring that strong confidence signals are not lost.

\section{Experimental Results}
\paragraph{Decomposition and Fusion Strategies}
To determine the optimal query decomposition and fusion pipeline, we conduct a study evaluating three proposed decomposition strategies alongside two fusion methods. Specifically, we randomly sample 30\% of the data across all datasets in the MultiConIR benchmark and evaluate each configuration on both the retrieval and reranking stages. As a reference, we first measure the performance of the undecomposed baseline. We observe that queries containing 2 or 3 conditions achieve higher scores than queries with more conditions. Motivated by this, we set $k = 2$ and $j = 3$ for fixed-size decomposition. Accordingly, queries with three or fewer conditions (Q1--Q3) already fall within this range and are left undecomposed. For Adaptive Decomposition, we let Qwen3-8B dynamically determine the optimal granularity for each query; the detailed prompt is provided in Appendix~\ref{app:prompt-qwen}.

\begin{table*}[!h]
\centering
\small
\setlength{\tabcolsep}{8pt}
\resizebox{\textwidth}{!}{ 
\begin{tabular}{lccccccccccc}
\toprule
Model & Q1 & Q2 & Q3 & Q4 & Q5 & Q6 & Q7 & Q8 & Q9 & Q10 & Avg \\
\midrule
\multicolumn{12}{c}{\textbf{NDCG@10}} \\
\midrule
\rowcolor{gray!10}
Base\textsuperscript{$\dagger$}
& 67.7 & 84.3 & 86.6
& 85.4 & 82.7 & 78.4 & 73.6 & 66.7 & 58.2 & 43.1 & 72.7 \\
\midrule
\rowcolor{gray!10}
Naive-sum
& 67.7 & 75.3 & 76.1
& 67.6 & 64.8 & 59.8 & 56.0 & 50.5 & 40.5 & 28.8 & 58.7 \\
\rowcolor{gray!10}
Naive-rrf
& 67.7 & 70.1 & 67.5
& 62.7 & 59.9 & 55.1 & 51.9 & 46.2 & 37.7 & 27.6 & 54.6 \\
\rowcolor{gray!10}
Adaptive-sum
& 67.7 & 78.9 & 81.8 
& 73.6 & 81.2 & 74.6 & 67.2 & 63.0 & \textbf{55.6} & 41.5 & 68.5 \\
\rowcolor{gray!10}
Adaptive-rrf
& 67.7 & 76.2 & 79.9 
& 68.1 & 79.5 & 71.0 & 61.9 & 59.6 & 53.1 & 39.5 & 65.7 \\
\rowcolor{gray!10}
Fixed-size-sum
& 67.7 & \textbf{84.3} & \textbf{86.6} 
& \textbf{82.5} & \textbf{81.5} & \textbf{78.0} & \textbf{72.0} & \textbf{65.2} & 55.5 & \textbf{42.3} & \textbf{71.6} \\
\rowcolor{gray!10}
Fixed-size-rrf
& 67.7 & \textbf{84.3} & \textbf{86.6} 
& 78.3 & 79.6 & 75.0 & 68.6 & 62.6 & 54.0 & 39.9 & 69.7 \\
\bottomrule
\end{tabular}
}
\caption{NDCG@10 performance comparison in retrieval stage across different decomposition and fusion strategies and query lengths using Qwen3-Embedding-0.6B. \textsuperscript{$\dagger$}Base (retrieve without decomposition) is shown as a reference. Values are scaled by 100. The "X-Y" naming convention denotes the combination of decomposition strategy X (e.g., naive, adaptive, fixed-size) and fusion method Y (e.g., scoresum, rrf).}
\label{tab:abalation-result-retrieval}
\end{table*}

\begin{table*}[!h]
\centering
\small
\setlength{\tabcolsep}{8pt}
\resizebox{\textwidth}{!}{ 
\begin{tabular}{lccccccccccc}
\toprule
Model & Q1 & Q2 & Q3 & Q4 & Q5 & Q6 & Q7 & Q8 & Q9 & Q10 & Avg \\
\midrule
\multicolumn{12}{c}{\textbf{NDCG@10}} \\
\midrule
\rowcolor{gray!10}
Pure Rerank\textsuperscript{$\dagger$}
& 80.1 & 87.1 & 86.2 
& 81.9 & 76.9 & 72.4 & 67.4 & 57.0 & 50.9 & 40.4 & 70.0 \\
\midrule
\rowcolor{gray!10}
Naive-sum
& 80.1 & 86.9 & 86.4 
& 86.8 & 83.7 & 79.5 & 74.7 & 67.8 & 61.2 & 49.9 & 75.7 \\
\rowcolor{gray!10}
Naive-rrf
& 80.1 & 86.7 & 86.1 
& 85.6 & 81.6 & 77.7 & 71.7 & 66.3 & 59.4 & 49.5 & 74.5 \\
\rowcolor{gray!10}
Adaptive-sum
& 80.1 & 87.0 & \textbf{86.6} 
& 85.3 & \textbf{83.7} & \textbf{80.6} & 72.8 & \textbf{69.3} & 60.0 & 51.2 & 75.7 \\
\rowcolor{gray!10}
Adaptive-rrf
& 80.1 & 86.8 & 86.2 
& 85.5 & 83.3 & 79.0 & 73.2 & 67.9 & 59.0 & 51.4 & 75.2 \\
\rowcolor{gray!10}
Fixed-size-sum
& 80.1 & \textbf{87.1} & 86.2
& \textbf{87.3} & 83.5 & 79.3 & \textbf{76.3} & 68.8 & \textbf{61.6} & \textbf{52.1} & \textbf{76.2} \\
\rowcolor{gray!10}
Fixed-size-rrf
& 80.1 & \textbf{87.1} & 86.2
& 86.1 & 83.5 & 78.6 & 74.2 & 66.8 & 59.0 & 52.0 & 75.4 \\
\bottomrule
\end{tabular}
}
\caption{NDCG@10 performance comparison in reranking stage across different decomposition and fusion strategies and query lengths using Qwen3-Reranker-0.6B. \textsuperscript{$\dagger$}Pure Rerank (rerank without decomposition) is shown as a reference. Values are scaled by 100. The "X-Y" naming convention same as Table \ref{tab:abalation-result-retrieval}.}
\label{tab:abalation-result-reranking}
\end{table*}

As shown in Tables~\ref{tab:abalation-result-retrieval} and \ref{tab:abalation-result-reranking}, fixed-size decomposition consistently yields the best overall performance across both the retrieval and reranking stages. Furthermore, the Score-Sum fusion method is shown to consistently outperform RRF. We therefore adopt these optimal configurations---fixed-size decomposition with $k=2$ and $j=3$, coupled with Score-Sum fusion---for all subsequent experiments. Beyond this direct conclusion, this study also reveals a striking stage-dependent pattern: query decomposition is detrimental at the retrieval stage but substantially boosts performance at the reranking stage. This observation motivates our Stage-Aware Decomposition Framework, illustrated in Figure~\ref{fig:rerank_pipeline}. We next verify whether the framework remains effective across the full benchmark and different retrieval and reranking models.

\paragraph{Retrieval Result}

\begin{table*}[!h]
\centering
\small
\setlength{\tabcolsep}{4pt}
\resizebox{\textwidth}{!}{ 
\begin{tabular}{lccccccccccc}
\toprule
Method & Q1 & Q2 & Q3 & Q4 & Q5 &
Q6 & Q7 & Q8 & Q9 & Q10 & Avg.(Q4--Q10) \\
\midrule
\multicolumn{12}{c}{\textbf{BGE-large-en-v1.5}} \\
\midrule
\rowcolor{gray!10}
Base (N@10)
& 69.8 & 80.7 & 84.7 & 84.2 & 81.5
& 77.1 & 72.0 & 64.8 & 55.8 & 42.3 & \textbf{68.2} \\
\rowcolor{gray!10}
Decomp-Merge (N@10)
& 69.8 & 80.7 & 84.7 & 78.0 & 77.5
& 74.9 & 65.8 & 60.7 & 52.6 & 38.0 & 63.9 \\
\rowcolor{gray!10}
Base (R@50)
& 85.8 & 94.5 & 97.7 & 98.7 & 99.5
& 99.5 & 99.7 & 99.5 & 99.6 & 99.8 & \textbf{99.5} \\
\rowcolor{gray!10}
Decomp-Merge (R@50)
& 85.8 & 94.5 & 97.7 & 95.3 & 97.3
& 98.4 & 95.5 & 96.9 & 97.7 & 94.1 & 96.5 \\
\midrule
\multicolumn{12}{c}{\textbf{Qwen3-Embedding-0.6B}} \\
\midrule
\rowcolor{gray!10}
Base (N@10)
& 67.7 & 84.0 & 86.2 & 85.8 & 82.5
& 78.7 & 73.4 & 66.5 & 58.5 & 45.5 & \textbf{70.1} \\
\rowcolor{gray!10}
\rowcolor{gray!10}
Decomp-Merge (N@10)
& 67.7 & 84.0 & 86.2 & 82.4 & 80.3
& 77.1 & 70.6 & 65.1 & 56.7 & 42.5 & 67.8 \\
\rowcolor{gray!10}
Base (R@50)
& 85.7 & 96.5 & 98.6 & 99.4 & 99.7
& 99.9 & 100.0 & 100.0 & 100.0 & 100.0 & \textbf{99.9} \\
\rowcolor{gray!10}
\rowcolor{gray!10}
Decomp-Merge (R@50)
& 85.7 & 96.5 & 98.6 & 97.5 & 98.6
& 99.2 & 98.7 & 99.5 & 99.6 & 99.0 & 98.9 \\
\midrule
\multicolumn{12}{c}{\textbf{Qwen3-Embedding-4B}} \\
\midrule
\rowcolor{gray!10}
Base (N@10)
& 69.8 & 83.7 & 86.9 & 86.0 & 82.5 
& 78.2 & 72.7 & 65.9 & 56.7 & 42.9 & \textbf{69.3} \\
\rowcolor{gray!10}
\rowcolor{gray!10}
Decomp-Merge (N@10)
& 69.8 & 83.7 & 86.9 & 83.3 & 80.7 
& 76.9 & 70.1 & 64.2 & 55.8 & 43.1 & 67.7 \\
\rowcolor{gray!10}
Base (R@50)
& 86.1 & 95.7 & 98.7 & 99.3 & 99.9 
& 99.8 & 100.0 & 100.0 & 100.0 & 100.0 & \textbf{99.9} \\
\rowcolor{gray!10}
\rowcolor{gray!10}
Decomp-Merge (R@50)
& 86.1 & 95.7 & 98.7 & 97.3 & 98.5 
& 99.3 & 98.6 & 99.4 & 99.5 & 99.3 & 98.8 \\
\midrule
\multicolumn{12}{c}{\textbf{Qwen3-Embedding-8B}} \\
\midrule
\rowcolor{gray!10}
Base (N@10)
& 72.6 & 85.1 & 87.4 & 86.3 & 83.5 
& 79.3 & 73.5 & 67.0 & 58.6 & 45.5 & \textbf{70.5} \\
\rowcolor{gray!10}
\rowcolor{gray!10}
Decomp-Merge (N@10)
& 72.6 & 85.1 & 87.4 & 83.4 & 81.0 
& 77.6 & 70.7 & 64.3 & 57.0 & 44.3 & 68.3 \\
\rowcolor{gray!10}
Base (R@50)
& 87.7 & 96.6 & 98.9 & 99.6 & 99.9 
& 100.0 & 100.0 & 100.0 & 100.0 & 100.0 & \textbf{99.9} \\
\rowcolor{gray!10}
\rowcolor{gray!10}
Decomp-Merge (R@50)
& 87.7 & 96.6 & 98.9 & 97.6 & 98.9 
& 99.5 & 98.7 & 99.4 & 99.5 & 99.4 & 99.0 \\
\bottomrule
\end{tabular}
}
\caption{Impact of query decomposition on retrieval across different query complexities. Values are scaled by 100. Base represents retrieve without decomposition. We provide full result in Appendix~\ref{sec:appendix}.}
\label{tab:main-result-retrieval}
\end{table*}

\begin{table*}[h]
\centering
\small
\setlength{\tabcolsep}{10pt}
\resizebox{\textwidth}{!}{ 
\begin{tabular}{lccccccccccc}
\toprule
Method & Q1 & Q2 & Q3 & Q4 & Q5 &
Q6 & Q7 & Q8 & Q9 & Q10 & Avg.(Q4--Q10) \\
\midrule

\multicolumn{12}{c}{\textbf{Qwen3-Reranker-0.6B}} \\
\midrule
\rowcolor{gray!10}
Orig. Base
& 69.8 & 80.7 & 84.7 & 84.2 & 81.5
& 77.1 & 72.0 & 64.8 & 55.8 & 42.3 & 68.2 \\
\rowcolor{gray!10}
Pure Rerank
& 79.6 & 86.9 & 86.4 & 82.3 & 77.8
& 72.0 & 67.7 & 58.1 & 50.8 & 40.2 & 64.1 \\
\rowcolor{gray!10}
Decomp-Merge
& 79.6 & 86.9 & 86.4 & 87.4 & 83.5
& 78.2 & 75.4 & 67.4 & 59.3 & 50.8 & 71.7 \\
\rowcolor{gray!10}
$\Delta$Decomp-Merge
& 0.0 & 0.0 & 0.0 & $\uparrow$5.1 & $\uparrow$5.7
& $\uparrow$6.2 & $\uparrow$7.7 & $\uparrow$9.3 & $\uparrow$8.5 & $\uparrow$10.6 & $\uparrow$7.6 \\

\midrule
\multicolumn{12}{c}{\textbf{bge-reranker-v2-m3}} \\
\midrule
\rowcolor{gray!10}
Orig. Base
& 69.8 & 80.7 & 84.7 & 84.2 & 81.5
& 77.1 & 72.0 & 64.8 & 55.8 & 42.3 & 68.2 \\
\rowcolor{gray!10}
Pure Rerank
& 79.4 & 89.0 & 90.1 & 88.2 & 84.8
& 80.6 & 74.4 & 67.3 & 57.9 & 43.9 & 71.0 \\
\rowcolor{gray!10}
Decomp-Merge
& 79.4 & 89.0 & 90.1 & 89.5 & 86.8
& 82.4 & 78.3 & 71.7 & 63.6 & 51.7 & 74.9 \\
\rowcolor{gray!10}
$\Delta$Decomp-Merge
& 0.0 & 0.0 & 0.0 & $\uparrow$1.3 & $\uparrow$2.0
& $\uparrow$1.8 & $\uparrow$3.9 & $\uparrow$4.4 & $\uparrow$5.7 & $\uparrow$7.8
& $\uparrow$3.9 \\

\midrule
\multicolumn{12}{c}{\textbf{Qwen3-Reranker-4B}} \\
\midrule
\rowcolor{gray!10}
Orig. Base
& 69.8 & 80.7 & 84.7 & 84.2 & 81.5
& 77.1 & 72.0 & 64.8 & 55.8 & 42.3 & 68.2 \\
\rowcolor{gray!10}
Pure Rerank
& 81.6 & 89.5 & 89.8 & 86.3 & 81.7
& 76.7 & 70.4 & 63.2 & 55.5 & 46.3 & 68.6 \\
\rowcolor{gray!10}
Decomp-Merge
& 81.6 & 89.5 & 89.8 & 89.6 & 86.2
& 81.5 & 77.7 & 71.6 & 64.2 & 57.1 & 75.4 \\
\rowcolor{gray!10}
$\Delta$Decomp-Merge
& 0.0 & 0.0 & 0.0 & $\uparrow$3.3 & $\uparrow$4.5
& $\uparrow$4.8 & $\uparrow$7.3 & $\uparrow$8.4 & $\uparrow$8.7 & $\uparrow$10.8 & $\uparrow$6.8 \\

\midrule
\multicolumn{12}{c}{\textbf{Qwen3-Reranker-8B}} \\
\midrule
\rowcolor{gray!10}
Orig. Base
& 69.8 & 80.7 & 84.7 & 84.2 & 81.5
& 77.1 & 72.0 & 64.8 & 55.8 & 42.3 & 68.2 \\
\rowcolor{gray!10}
Pure Rerank
& 81.3 & 90.8 & 92.1 & 90.3 & 86.9
& 82.1 & 76.5 & 68.8 & 60.8 & 51.9 & 73.9 \\
\rowcolor{gray!10}
Decomp-Merge
& 81.3 & 90.8 & 92.1 & 91.6 & 88.4
& 85.2 & 81.2 & 75.7 & 70.0 & 66.6 & 79.8 \\
\rowcolor{gray!10}
$\Delta$Decomp-Merge
& 0.0 & 0.0 & 0.0 & $\uparrow$1.3 & $\uparrow$1.5
& $\uparrow$3.1 & $\uparrow$4.7 & $\uparrow$6.9 & $\uparrow$9.2 & $\uparrow$14.7 & $\uparrow$5.9 \\

\bottomrule
\end{tabular}
}
\caption{Impact of query decomposition on reranking across different query complexities. Values are scaled by 100. Orig. Base: retrieval-only baseline; Pure Rerank: rerank without decomposition; Decomp-Merge: rerank each sub-query then merge; $\Delta$Decomp-Merge: improvement over Pure Rerank. We provide full results in Appendix~\ref{sec:appendix}.}
\label{tab:main-result-reranking}
\end{table*}

As shown in Table~\ref{tab:main-result-retrieval}, applying query decomposition to complex queries (Queries 4--10) consistently degrades performance across all evaluated models in terms of both NDCG@10 and Recall@50. Specifically, for the NDCG@10 metric, all models experience a consistent drop following decomposition. The most severe decline is observed with BGE-large-en-v1.5, which exhibits a 4.3-point degradation, while the Qwen series models show drops ranging from 1.6 to 2.2 points. Similarly, for Recall@50, both query decomposition strategies consistently underperform the standard baseline, registering an absolute reduction of approximately 1 to 3 points. These findings confirm that query decomposition fundamentally impairs performance at the initial retrieval stage.

\paragraph{Reranking Result}
As Table~\ref{tab:main-result-reranking} indicates, standard reranking does not guarantee improvements and can even degrade retrieval quality. For instance, Qwen3-Reranker-0.6B suffers a 4.1-point drop on complex queries (Queries 4--10) compared to the unreranked baseline. Conversely, integrating query decomposition into the reranking stage yields substantial gains, outperforming standard reranking by 7.6 points and successfully surpassing the baseline. Notably, this performance gain over the baseline scales robustly with model size: the larger Qwen3-Reranker-8B achieves an 11.6-point improvement over the baseline and a 5.9-point gain over standard reranking. This result also aligns with our initial hypothesis, confirming that query decomposition is beneficial at the reranking stage.

\section{Analysis}
\subsection{Why Retrieval Fails}


\begin{figure}[t]
    \centering
    \includegraphics[width=\linewidth]{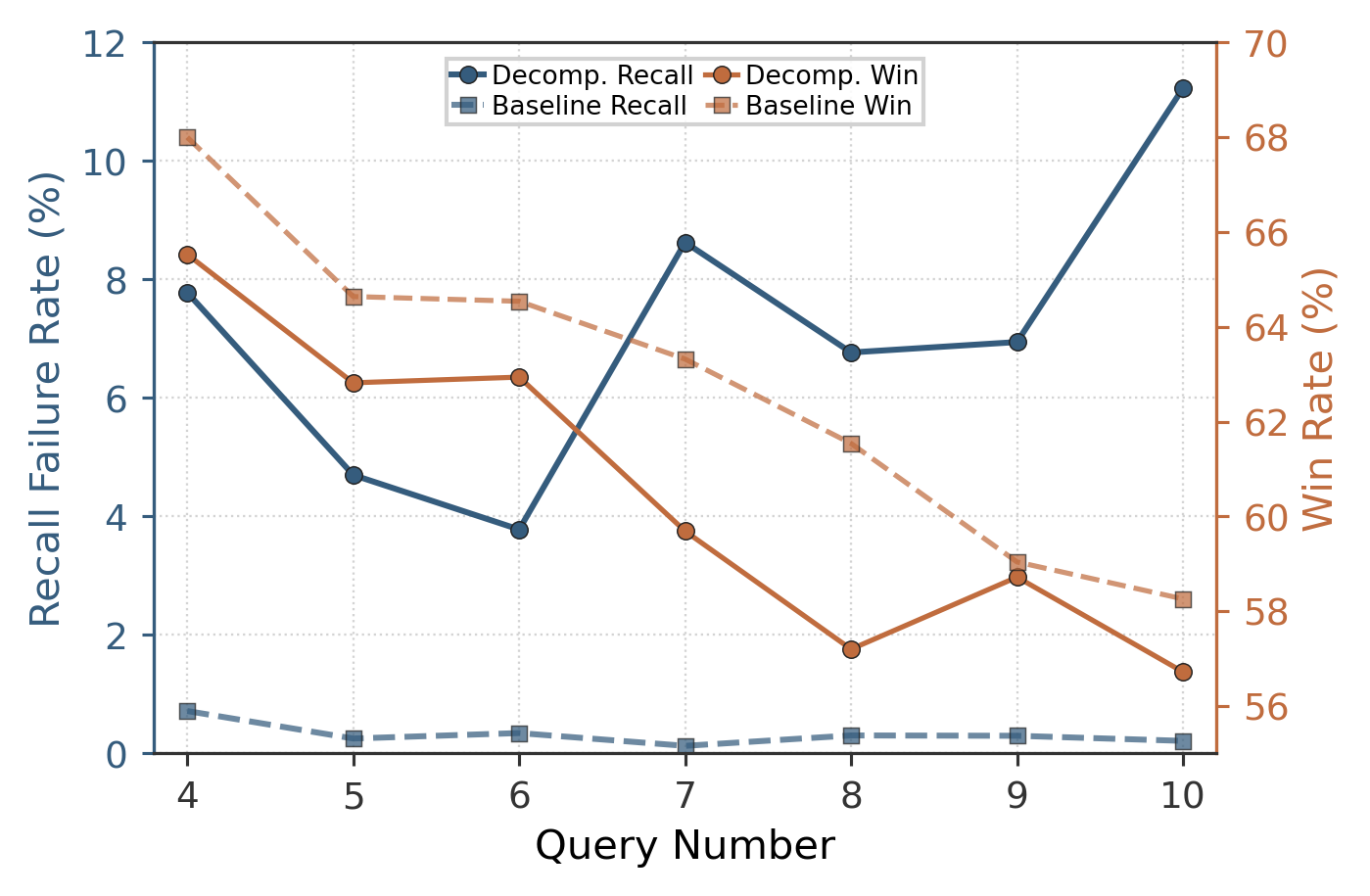}
    \caption{
    Recall failure rate (\%) and average win rate of baseline queries versus decomposed sub-queries in retrieval. Undecomposed queries consistently exhibit a lower recall failure  than sub-queries, while baseline retrieval achieves a higher win rate than retrieval with query decomposition.
    }
    \label{fig:recall_failure}

\end{figure}

\paragraph{High Incidence of Recall Failure in Sub-queries}
Query decomposition degrades initial retrieval primarily by missing target documents entirely. We define this as \textit{recall failure}: failing to retrieve any positive document within the top 50 candidates. As Figure~\ref{fig:recall_failure} illustrates, baseline queries maintain a recall failure rate strictly below 1\%, whereas decomposed sub-queries experience a dramatic spike to 3.5\%–12\%. This drop in foundational recall establishes a hard bottleneck, explaining why early decomposition is detrimental to the overall pipeline.

\paragraph{Increased Susceptibility to Easy Negatives}
Beyond recall failure, applying decomposition during the initial retrieval phase also makes the system more susceptible to easy negative (EN) documents. As shown in Table~\ref{tab:en_percentage}, decomposed queries consistently retrieve more ENs within their top 50 candidates compared to the baseline. We provide an example in Appendix~\ref{app:recall_failure} illustrating how semantic dilution in a decomposed sub-query leads to the retrieval of easy negative documents.

\begin{table}[h]
\centering
\small
\setlength{\tabcolsep}{5pt}

\begin{tabular}{lccc}
\toprule
Dataset & Baseline & Decompose & $\Delta$ \\
\midrule
Books            & 78.88 & 80.81 & +1.93 \\
People           & 78.18 & 78.39 & +0.21 \\
Movies           & 79.14 & 80.46 & +1.32 \\
Medical Cases    & 78.59 & 80.17 & +1.58 \\
Legal Documents  & 78.33 & 80.04 & +1.71 \\
\midrule
Average          & 78.62 & 79.97 & +1.35 \\
\bottomrule
\end{tabular}

\caption{
Probability (\%) of retrieving easy negative documents within the top 50 candidates for baseline retrieval and retrieval with query decomposition.
}
\label{tab:en_percentage}

\end{table}

\paragraph{Diminished Fine-Grained Capacity}
We measure fine-grained discriminative capacity using the win rate metric to evaluate whether decomposition improves the relative ranking among highly similar items. As shown in Figure~\ref{fig:recall_failure}, baseline retrieval consistently outperforms the decomposed approach across all query lengths. While capacity declines for both methods as queries lengthen, decomposition provides no fine-grained advantage, maintaining a strictly lower win rate throughout.

The concurrent rise in recall failures and easy negative (EN) retrieval highlights a fundamental flaw in early query decomposition. Fragmenting a cohesive, information-rich query into isolated sub-queries dilutes its overall semantic constraints. In a large corpus, these weakened constraints not only cause the model to miss target documents but also allow it to be misled by ENs. Furthermore, the win rate analysis confirms that decomposition offers no compensatory benefits; it fails to enhance fine-grained discriminative capacity, leaving the model unable to distinguish between similar candidates.


\begin{figure}[!t]
    \centering
    \includegraphics[width=0.48\textwidth]{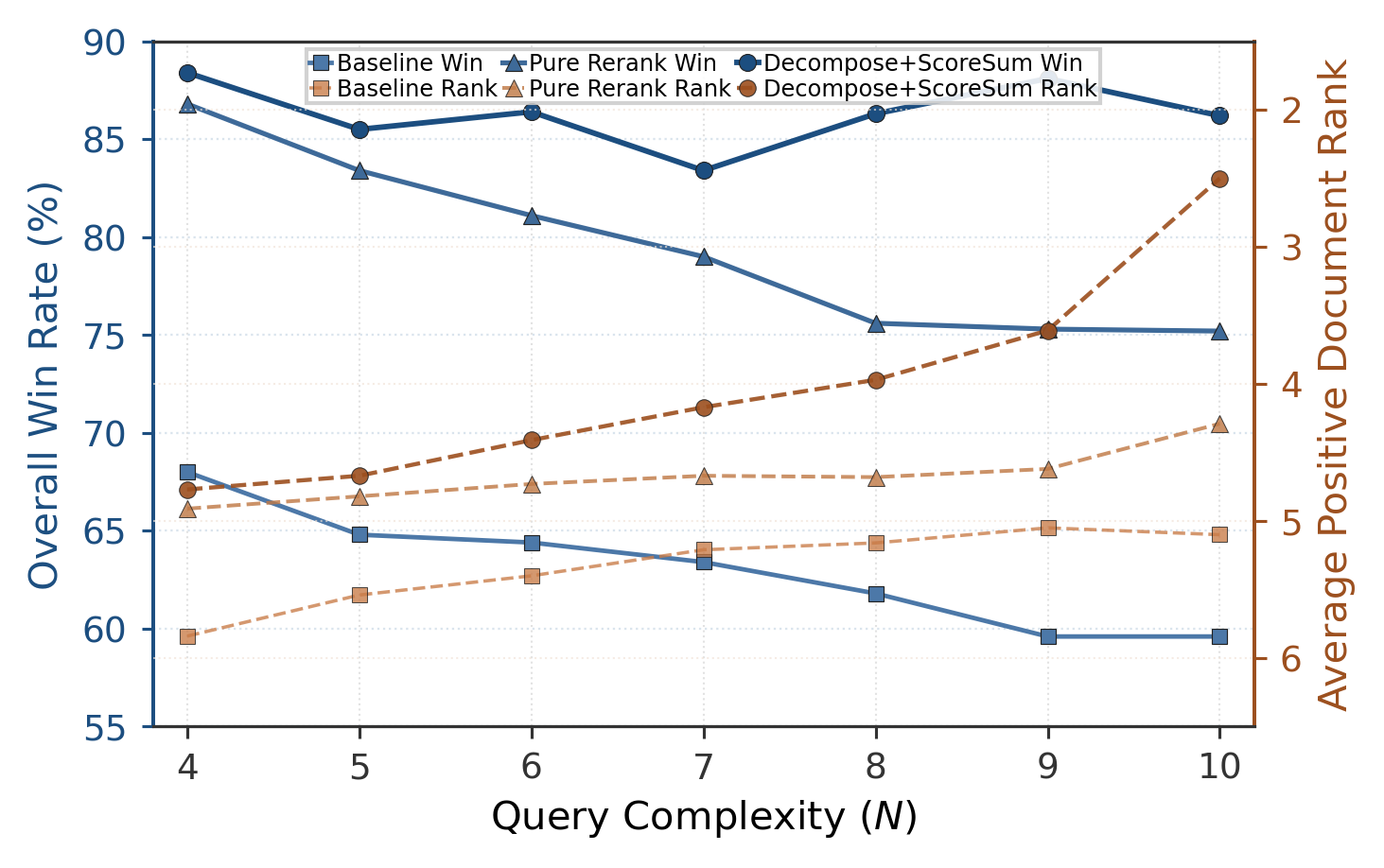}
    \caption{
    Average win rate (\%) and average rank position of positive documents for baseline reranking and reranking with query decomposition across different query complexity levels. Decomposition in reranking achieves a higher win rate and a better (lower) average rank of positive documents than both pure reranking and baseline retrieval.
    }
    \label{fig:win_rate_positive_document_reranking}
\end{figure}

\subsection{Why Reranking Works}
\paragraph{Enhanced Fine-Grained Capacity}
We further evaluate fine-grained discriminative capacity using the win rate metric. As Figure~\ref{fig:win_rate_positive_document_reranking} indicates, decomposition method significantly outperform standard reranking. Notably, this performance gap widens for highly complex, multi-condition queries. This demonstrates that query decomposition effectively isolates specific constraints, enabling the model to better capture and distinguish fine-grained details.


\paragraph{Elevated Positive Document Ranking}
Given that initial recall is already high for complex queries (Queries 4--10), the primary objective of the reranker is to elevate true positives above hard negatives. To evaluate this capability, we measure the average rank position of positive documents. As shown in Figure~\ref{fig:win_rate_positive_document_reranking}, applying query decomposition during reranking places positive documents significantly higher than standard reranking. This demonstrates that decomposition empowers the model to better differentiate true positives from hard negatives, thereby improving overall ranking quality.


The concurrent improvements in win rate and positive document ranking confirm the efficacy of query decomposition in reranking. By fragmenting complex queries, the model is forced to attend to individual semantic constraints, significantly enhancing its fine-grained discriminative capacity. Furthermore, because reranking operates on a constrained candidate pool, it avoids the context dilution inherent to large-scale retrieval. This focused environment makes decomposition a strictly beneficial strategy for final document scoring.




\subsection{Generalization Analysis on SSRB}
To demonstrate the generalizability of our end-to-end pipeline, we extend our evaluation to the Semi-Structured Retrieval Benchmark (SSRB)~\cite{zhang2025ssrb}, a widely adopted dataset for multi-condition information retrieval. We randomly select three representative schemas for this analysis.

To ensure high-quality fixed-size query decomposition, we utilize GPT-5.3~\cite{DBLP:journals/corr/abs-2601-03267} with tailored prompts (detailed in Appendix~\ref{app:prompt}). For the underlying models, we deploy Qwen3-Embedding-4B for retrieval and Qwen3-Reranker-8B for reranking. In our initial experiments, applying decomposition directly during the retrieval phase resulted in a sharp performance drop compared to the baseline. However, by deploying our proposed end-to-end pipeline, we observe substantial performance improvements. Detailed results are presented in Table~\ref{tab:ssrb_results}.

\begin{table}[h]
\centering
\small
\setlength{\tabcolsep}{5pt}
\resizebox{\linewidth}{!}{
\begin{tabular}{llccc}
\toprule
\multirow{2}{*}{\textbf{Stage}} & \multirow{2}{*}{\textbf{Setting}} & \multicolumn{3}{c}{\textbf{Schema}} \\
\cmidrule{3-5}
 & & \textbf{Camera} & \textbf{Employee} & \textbf{API} \\
\midrule
\multicolumn{5}{c}{\textbf{NDCG@10}} \\
\midrule
\multirow{2}{*}{Retrieval}
 & Base                   & \textbf{11.1} & \textbf{10.0} & \textbf{19.3} \\
 & Decomp                 & 8.9           & 6.8           & 12.6 \\
\midrule
\multirow{2}{*}{Reranking}
 & Normal Rerank          & 13.2          & 10.6          & 10.3 \\
 & Stage-Aware            & \textbf{14.9} & \textbf{12.7} & \textbf{11.4} \\
\midrule
\multicolumn{5}{c}{\textbf{Recall@100}} \\
\midrule
\multirow{2}{*}{Retrieval}
 & Base                   & \textbf{20.3} & \textbf{21.4} & \textbf{26.8} \\
 & Decomp                 & 14.7          & 18.4          & 17.1 \\
\bottomrule
\end{tabular}
}
\caption{Impact of query decomposition on retrieval and reranking performance across selected SSRB schemas. Values are scaled by 100.}
\label{tab:ssrb_results}
\end{table}

\subsection{Computational Efficiency Analysis}

To quantify the computational overhead introduced by query decomposition, we evaluate the inference efficiency during the reranking stage. We compare the No-Decompose baseline with decomposition-based reranking across four reranker models: \textbf{bge-reranker-v2-m3}, \textbf{Qwen3-Reranker-0.6B}, \textbf{Qwen3-Reranker-4B}, and \textbf{Qwen3-Reranker-8B}, using the Books dataset.

For each query, we measure the total time required to rerank its 50 candidate documents. Under the decomposition setting, the reported runtime includes scoring all sub-query--candidate pairs and aggregating their scores. We use vLLM for inference acceleration with a batch size of 50. All experiments are conducted on a single NVIDIA A800 80GB GPU.

\begin{table}[h]
\centering
\setlength{\tabcolsep}{3pt}
\renewcommand{\arraystretch}{1.1}

\begin{tabular}{lccc}
\toprule
\textbf{Model} &
\textbf{No-Decomp.} &
\textbf{Decomp.} &
\textbf{Ratio} \\
\midrule
Qwen3-0.6B & 0.2202 & 0.4693 & 2.13$\times$ \\
BGE-v2-m3  & 0.1706 & 0.3439 & 2.02$\times$ \\
Qwen3-4B   & 0.6001 & 1.2969 & 2.16$\times$ \\
Qwen3-8B   & 0.4536 & 0.9776 & 2.16$\times$ \\
\bottomrule
\end{tabular}

\caption{Inference efficiency of decomposition-based reranking on the Books dataset. Runtime is reported in seconds per query (s/q), where each query contains 50 candidate documents.}
\label{tab:efficiency}
\end{table}

\section{Conclusion}
In this work, we conducted a systematic empirical study on the role of query decomposition in multi-condition information retrieval. Our findings reveal a critical stage-dependency: applying decomposition during the initial retrieval phase degrades performance by inducing severe recall failure and introducing easy negatives. Conversely, deploying decomposition exclusively during the reranking phase significantly enhances the model's fine-grained discriminative capacity against hard negatives. 
Based on these insights, we proposed a novel \textbf{Stage-Aware Decomposition Framework}. By retaining the monolithic query for robust candidate generation and leveraging strategically partitioned sub-queries for cross-encoder reranking, our framework effectively balances global semantic context with strict logical verification. Ultimately, this approach offers a principled and highly effective pipeline for modern retrieval systems to handle complex, compositional user intents.

\section*{Limitations}
First, while our study explores three representative query decomposition strategies and two merging techniques across distinct multi-condition retrieval benchmarks, the combinatorial design space of multi-stage pipelines is vast. Future work could investigate alternative decomposition granularities along with various fusion methods to further explore the performance boundaries and trade-offs of stage-aware retrieval. Second, to establish a highly controlled and rigorous experimental baseline, our evaluation primarily focuses on the Qwen3 model family across both pipeline stages. While these consistent model configurations provide clean, unconfounded insights into stage-dependent behavior, exploring the cross-architecture generalizability of our findings to other emerging dense retrievers and rerankers represents a promising direction for broader validation. 
Third, this work is conceptually dedicated to isolating and diagnosing the intrinsic mechanics of query decomposition across pipeline stages, rather than focusing on engineering-driven performance competition with miscellaneous retrieval-augmentation frameworks. Extending this stage-aware principle to integrate with other advanced indexing or prompting paradigms remains an open avenue for future exploration.

\bibliography{custom}

@inproceedings{DBLP:conf/chi/YeeSLH03,
  author       = {Ka{-}Ping Yee and
                  Kirsten Swearingen and
                  Kevin Li and
                  Marti A. Hearst},
  editor       = {Gilbert Cockton and
                  Panu Korhonen},
  title        = {Faceted metadata for image search and browsing},
  booktitle    = {Proceedings of the 2003 Conference on Human Factors in Computing Systems,
                  {CHI} 2003, Ft. Lauderdale, Florida, USA, April 5-10, 2003},
  pages        = {401--408},
  publisher    = {{ACM}},
  year         = {2003},
  url          = {https://doi.org/10.1145/642611.642681},
  doi          = {10.1145/642611.642681},
  bibsource    = {dblp computer science bibliography, https://dblp.org}
}

@misc{li2023making,
  title={Making Large Language Models A Better Foundation For Dense Retrieval},
  author={Chaofan Li and Zheng Liu and Shitao Xiao and Yingxia Shao},
  year={2023},
  eprint={2312.15503},
  archivePrefix={arXiv},
  primaryClass={cs.CL}
}

@misc{chen2024bge,
  title={BGE M3-Embedding: Multi-Lingual, Multi-Functionality, Multi-Granularity Text Embeddings Through Self-Knowledge Distillation},
  author={Jianlv Chen and Shitao Xiao and Peitian Zhang and Kun Luo and Defu Lian and Zheng Liu},
  year={2024},
  eprint={2402.03216},
  archivePrefix={arXiv},
  primaryClass={cs.CL}
}

@inproceedings{DBLP:conf/emnlp/PressZMSSL23,
  author       = {Ofir Press and
                  Muru Zhang and
                  Sewon Min and
                  Ludwig Schmidt and
                  Noah A. Smith and
                  Mike Lewis},
  editor       = {Houda Bouamor and
                  Juan Pino and
                  Kalika Bali},
  title        = {Measuring and Narrowing the Compositionality Gap in Language Models},
  booktitle    = {Findings of the Association for Computational Linguistics: {EMNLP}
                  2023, Singapore, December 6-10, 2023},
  series       = {Findings of {ACL}},
  pages        = {5687--5711},
  publisher    = {Association for Computational Linguistics},
  year         = {2023},
  url          = {https://doi.org/10.18653/v1/2023.findings-emnlp.378},
  doi          = {10.18653/V1/2023.FINDINGS-EMNLP.378},
  bibsource    = {dblp computer science bibliography, https://dblp.org}
}

@inproceedings{DBLP:conf/emnlp/KarpukhinOMLWEC20,
  author       = {Vladimir Karpukhin and
                  Barlas Oguz and
                  Sewon Min and
                  Patrick Lewis and
                  Ledell Wu and
                  Sergey Edunov and
                  Danqi Chen and
                  Wen{-}tau Yih},
  editor       = {Bonnie Webber and
                  Trevor Cohn and
                  Yulan He and
                  Yang Liu},
  title        = {Dense Passage Retrieval for Open-Domain Question Answering},
  booktitle    = {Proceedings of the 2020 Conference on Empirical Methods in Natural
                  Language Processing, {EMNLP} 2020, Online, November 16-20, 2020},
  pages        = {6769--6781},
  publisher    = {Association for Computational Linguistics},
  year         = {2020},
  url          = {https://doi.org/10.18653/v1/2020.emnlp-main.550},
  doi          = {10.18653/V1/2020.EMNLP-MAIN.550},
  bibsource    = {dblp computer science bibliography, https://dblp.org}
}

@article{DBLP:journals/ftir/RobertsonZ09,
  author       = {Stephen E. Robertson and
                  Hugo Zaragoza},
  title        = {The Probabilistic Relevance Framework: {BM25} and Beyond},
  journal      = {Found. Trends Inf. Retr.},
  volume       = {3},
  number       = {4},
  pages        = {333--389},
  year         = {2009},
  url          = {https://doi.org/10.1561/1500000019},
  doi          = {10.1561/1500000019},
  bibsource    = {dblp computer science bibliography, https://dblp.org}
}

@inproceedings{DBLP:conf/naacl/DevlinCLT19,
  author       = {Jacob Devlin and
                  Ming{-}Wei Chang and
                  Kenton Lee and
                  Kristina Toutanova},
  editor       = {Jill Burstein and
                  Christy Doran and
                  Thamar Solorio},
  title        = {{BERT:} Pre-training of Deep Bidirectional Transformers for Language
                  Understanding},
  booktitle    = {Proceedings of the 2019 Conference of the North American Chapter of
                  the Association for Computational Linguistics: Human Language Technologies,
                  {NAACL-HLT} 2019, Minneapolis, MN, USA, June 2-7, 2019, Volume 1 (Long
                  and Short Papers)},
  pages        = {4171--4186},
  publisher    = {Association for Computational Linguistics},
  year         = {2019},
  url          = {https://doi.org/10.18653/v1/n19-1423},
  doi          = {10.18653/V1/N19-1423},
  bibsource    = {dblp computer science bibliography, https://dblp.org}
}

@inproceedings{DBLP:conf/emnlp/0001YMWRCYR23,
  author       = {Weiwei Sun and
                  Lingyong Yan and
                  Xinyu Ma and
                  Shuaiqiang Wang and
                  Pengjie Ren and
                  Zhumin Chen and
                  Dawei Yin and
                  Zhaochun Ren},
  editor       = {Houda Bouamor and
                  Juan Pino and
                  Kalika Bali},
  title        = {Is ChatGPT Good at Search? Investigating Large Language Models as
                  Re-Ranking Agents},
  booktitle    = {Proceedings of the 2023 Conference on Empirical Methods in Natural
                  Language Processing, {EMNLP} 2023, Singapore, December 6-10, 2023},
  pages        = {14918--14937},
  publisher    = {Association for Computational Linguistics},
  year         = {2023},
  url          = {https://doi.org/10.18653/v1/2023.emnlp-main.923},
  doi          = {10.18653/V1/2023.EMNLP-MAIN.923},
  bibsource    = {dblp computer science bibliography, https://dblp.org}
}

@inproceedings{DBLP:conf/acl/GaoMLC23,
  author       = {Luyu Gao and
                  Xueguang Ma and
                  Jimmy Lin and
                  Jamie Callan},
  editor       = {Anna Rogers and
                  Jordan L. Boyd{-}Graber and
                  Naoaki Okazaki},
  title        = {Precise Zero-Shot Dense Retrieval without Relevance Labels},
  booktitle    = {Proceedings of the 61st Annual Meeting of the Association for Computational
                  Linguistics (Volume 1: Long Papers), {ACL} 2023, Toronto, Canada,
                  July 9-14, 2023},
  pages        = {1762--1777},
  publisher    = {Association for Computational Linguistics},
  year         = {2023},
  url          = {https://doi.org/10.18653/v1/2023.acl-long.99},
  doi          = {10.18653/V1/2023.ACL-LONG.99},
  bibsource    = {dblp computer science bibliography, https://dblp.org}
}

@inproceedings{DBLP:conf/acl/MinZZH19,
  author       = {Sewon Min and
                  Victor Zhong and
                  Luke Zettlemoyer and
                  Hannaneh Hajishirzi},
  editor       = {Anna Korhonen and
                  David R. Traum and
                  Llu{\'{\i}}s M{\`{a}}rquez},
  title        = {Multi-hop Reading Comprehension through Question Decomposition and
                  Rescoring},
  booktitle    = {Proceedings of the 57th Conference of the Association for Computational
                  Linguistics, {ACL} 2019, Florence, Italy, July 28- August 2, 2019,
                  Volume 1: Long Papers},
  pages        = {6097--6109},
  publisher    = {Association for Computational Linguistics},
  year         = {2019},
  url          = {https://doi.org/10.18653/v1/p19-1613},
  doi          = {10.18653/V1/P19-1613},
  bibsource    = {dblp computer science bibliography, https://dblp.org}
}

@article{DBLP:journals/corr/abs-2503-09516,
  author       = {Bowen Jin and
                  Hansi Zeng and
                  Zhenrui Yue and
                  Dong Wang and
                  Hamed Zamani and
                  Jiawei Han},
  title        = {Search-R1: Training LLMs to Reason and Leverage Search Engines with
                  Reinforcement Learning},
  journal      = {CoRR},
  volume       = {abs/2503.09516},
  year         = {2025},
  url          = {https://doi.org/10.48550/arXiv.2503.09516},
  doi          = {10.48550/ARXIV.2503.09516},
  eprinttype   = {arXiv},
  eprint       = {2503.09516},
  bibsource    = {dblp computer science bibliography, https://dblp.org}
}

@article{DBLP:journals/corr/abs-2508-09303,
  author       = {Shu Zhao and
                  Tan Yu and
                  Anbang Xu and
                  Japinder Singh and
                  Aaditya Shukla and
                  Rama Akkiraju},
  title        = {ParallelSearch: Train your LLMs to Decompose Query and Search Sub-queries
                  in Parallel with Reinforcement Learning},
  journal      = {CoRR},
  volume       = {abs/2508.09303},
  year         = {2025},
  url          = {https://doi.org/10.48550/arXiv.2508.09303},
  doi          = {10.48550/ARXIV.2508.09303},
  eprinttype   = {arXiv},
  eprint       = {2508.09303},
  bibsource    = {dblp computer science bibliography, https://dblp.org}
}

@inproceedings{lu-etal-2025-multiconir,
    title = "{M}ulti{C}on{IR}: Towards Multi-Condition Information Retrieval",
    author = "Lu, Xuan  and
      Liu, Sifan  and
      Yin, Bochao  and
      Li, Yongqi  and
      Chen, Xinghao  and
      Su, Hui  and
      Jin, Yaohui  and
      Zeng, Wenjun  and
      Shen, Xiaoyu",
    editor = "Christodoulopoulos, Christos  and
      Chakraborty, Tanmoy  and
      Rose, Carolyn  and
      Peng, Violet",
    booktitle = "Findings of the Association for Computational Linguistics: EMNLP 2025",
    month = nov,
    year = "2025",
    address = "Suzhou, China",
    publisher = "Association for Computational Linguistics",
    url = "https://aclanthology.org/2025.findings-emnlp.726/",
    doi = "10.18653/v1/2025.findings-emnlp.726",
    pages = "13471--13494",
    ISBN = "979-8-89176-335-7"
}

@inproceedings{
zhang2025ssrb,
title={{SSRB}: Direct Natural Language Querying to Massive Heterogeneous Semi-Structured Data},
author={Xin Zhang and Mingxin Li and Yanzhao Zhang and Dingkun Long and Yongqi Li and Yinghui Li and Pengjun Xie and Meishan Zhang and Wenjie Li and Min Zhang and Philip S. Yu},
booktitle={The Thirty-ninth Annual Conference on Neural Information Processing Systems Datasets and Benchmarks Track},
year={2025},
url={https://openreview.net/forum?id=LuMGoG6lBA}
}

@inproceedings{DBLP:conf/sigir/CormackCB09,
  author       = {Gordon V. Cormack and
                  Charles L. A. Clarke and
                  Stefan B{\"{u}}ttcher},
  editor       = {James Allan and
                  Javed A. Aslam and
                  Mark Sanderson and
                  ChengXiang Zhai and
                  Justin Zobel},
  title        = {Reciprocal rank fusion outperforms condorcet and individual rank learning
                  methods},
  booktitle    = {Proceedings of the 32nd Annual International {ACM} {SIGIR} Conference
                  on Research and Development in Information Retrieval, {SIGIR} 2009,
                  Boston, MA, USA, July 19-23, 2009},
  pages        = {758--759},
  publisher    = {{ACM}},
  year         = {2009},
  url          = {https://doi.org/10.1145/1571941.1572114},
  doi          = {10.1145/1571941.1572114},
  bibsource    = {dblp computer science bibliography, https://dblp.org}
}

@inproceedings{DBLP:conf/nips/NguyenRSGTMD16,
  author       = {Tri Nguyen and
                  Mir Rosenberg and
                  Xia Song and
                  Jianfeng Gao and
                  Saurabh Tiwary and
                  Rangan Majumder and
                  Li Deng},
  editor       = {Tarek Richard Besold and
                  Antoine Bordes and
                  Artur S. d'Avila Garcez and
                  Greg Wayne},
  title        = {{MS} {MARCO:} {A} Human Generated MAchine Reading COmprehension Dataset},
  booktitle    = {Proceedings of the Workshop on Cognitive Computation: Integrating
                  neural and symbolic approaches 2016 co-located with the 30th Annual
                  Conference on Neural Information Processing Systems {(NIPS} 2016),
                  Barcelona, Spain, December 9, 2016},
  series       = {{CEUR} Workshop Proceedings},
  volume       = {1773},
  publisher    = {CEUR-WS.org},
  year         = {2016},
  url          = {https://ceur-ws.org/Vol-1773/CoCoNIPS\_2016\_paper9.pdf},
  bibsource    = {dblp computer science bibliography, https://dblp.org}
}

@article{DBLP:journals/corr/abs-2104-08663,
  author       = {Nandan Thakur and
                  Nils Reimers and
                  Andreas R{\"{u}}ckl{\'{e}} and
                  Abhishek Srivastava and
                  Iryna Gurevych},
  title        = {{BEIR:} {A} Heterogenous Benchmark for Zero-shot Evaluation of Information
                  Retrieval Models},
  journal      = {CoRR},
  volume       = {abs/2104.08663},
  year         = {2021},
  url          = {https://arxiv.org/abs/2104.08663},
  eprinttype   = {arXiv},
  eprint       = {2104.08663},
  bibsource    = {dblp computer science bibliography, https://dblp.org}
}

@inproceedings{DBLP:conf/iclr/SuYXSMWLSST0YA025,
  author       = {Hongjin Su and
                  Howard Yen and
                  Mengzhou Xia and
                  Weijia Shi and
                  Niklas Muennighoff and
                  Han{-}yu Wang and
                  Haisu Liu and
                  Quan Shi and
                  Zachary S. Siegel and
                  Michael Tang and
                  Ruoxi Sun and
                  Jinsung Yoon and
                  Sercan {\"{O}}. Arik and
                  Danqi Chen and
                  Tao Yu},
  title        = {{BRIGHT:} {A} Realistic and Challenging Benchmark for Reasoning-Intensive
                  Retrieval},
  booktitle    = {The Thirteenth International Conference on Learning Representations,
                  {ICLR} 2025, Singapore, April 24-28, 2025},
  publisher    = {OpenReview.net},
  year         = {2025},
  url          = {https://openreview.net/forum?id=ykuc5q381b},
  bibsource    = {dblp computer science bibliography, https://dblp.org}
}

@article{DBLP:journals/corr/abs-2402-14151,
  author       = {Xiaoyue Wang and
                  Jianyou Wang and
                  Weili Cao and
                  Kaicheng Wang and
                  Ramamohan Paturi and
                  Leon Bergen},
  title        = {{BIRCO:} {A} Benchmark of Information Retrieval Tasks with Complex
                  Objectives},
  journal      = {CoRR},
  volume       = {abs/2402.14151},
  year         = {2024},
  url          = {https://doi.org/10.48550/arXiv.2402.14151},
  doi          = {10.48550/ARXIV.2402.14151},
  eprinttype    = {arXiv},
  eprint       = {2402.14151},
  bibsource    = {dblp computer science bibliography, https://dblp.org}
}

@article{DBLP:journals/tois/LinYNTWL21,
  author       = {Sheng{-}Chieh Lin and
                  Jheng{-}Hong Yang and
                  Rodrigo Nogueira and
                  Ming{-}Feng Tsai and
                  Chuan{-}Ju Wang and
                  Jimmy Lin},
  title        = {Multi-Stage Conversational Passage Retrieval: An Approach to Fusing
                  Term Importance Estimation and Neural Query Rewriting},
  journal      = {{ACM} Trans. Inf. Syst.},
  volume       = {39},
  number       = {4},
  pages        = {48:1--48:29},
  year         = {2021},
  url          = {https://doi.org/10.1145/3446426},
  doi          = {10.1145/3446426},
  bibsource    = {dblp computer science bibliography, https://dblp.org}
}

@inproceedings{lin-etal-2023-decomposing,
    title = "Decomposing Complex Queries for Tip-of-the-tongue Retrieval",
    author = "Lin, Kevin  and
      Lo, Kyle  and
      Gonzalez, Joseph  and
      Klein, Dan",
    editor = "Bouamor, Houda  and
      Pino, Juan  and
      Bali, Kalika",
    booktitle = "Findings of the Association for Computational Linguistics: EMNLP 2023",
    month = dec,
    year = "2023",
    address = "Singapore",
    publisher = "Association for Computational Linguistics",
    url = "https://aclanthology.org/2023.findings-emnlp.367/",
    doi = "10.18653/v1/2023.findings-emnlp.367",
    pages = "5521--5533"
}

@inproceedings{ammann-etal-2025-question,
    title = "Question Decomposition for Retrieval-Augmented Generation",
    author = "Ammann, Paul J. L.  and
      Golde, Jonas  and
      Akbik, Alan",
    editor = "Zhao, Jin  and
      Wang, Mingyang  and
      Liu, Zhu",
    booktitle = "Proceedings of the 63rd Annual Meeting of the Association for Computational Linguistics (Volume 4: Student Research Workshop)",
    month = jul,
    year = "2025",
    address = "Vienna, Austria",
    publisher = "Association for Computational Linguistics",
    url = "https://aclanthology.org/2025.acl-srw.32/",
    doi = "10.18653/v1/2025.acl-srw.32",
    pages = "497--507",
    ISBN = "979-8-89176-254-1"
}

@inproceedings{DBLP:conf/icml/LiuLWC25,
  author       = {Yaoyang Liu and
                  Junlin Li and
                  Yinjun Wu and
                  Zhen Chen},
  editor       = {Aarti Singh and
                  Maryam Fazel and
                  Daniel Hsu and
                  Simon Lacoste{-}Julien and
                  Felix Berkenkamp and
                  Tegan Maharaj and
                  Kiri Wagstaff and
                  Jerry Zhu},
  title        = {{POQD:} Performance-Oriented Query Decomposer for Multi-vector retrieval},
  booktitle    = {Forty-second International Conference on Machine Learning, {ICML}
                  2025, Vancouver, BC, Canada, July 13-19, 2025},
  series       = {Proceedings of Machine Learning Research},
  volume       = {267},
  publisher    = {{PMLR} / OpenReview.net},
  year         = {2025},
  url          = {https://proceedings.mlr.press/v267/liu25ag.html},
  bibsource    = {dblp computer science bibliography, https://dblp.org}
}

@article{DBLP:journals/corr/abs-2510-18633,
  author       = {Roxana Petcu and
                  Kenton Murray and
                  Daniel Khashabi and
                  Evangelos Kanoulas and
                  Maarten de Rijke and
                  Dawn J. Lawrie and
                  Kevin Duh},
  title        = {Query Decomposition for {RAG:} Balancing Exploration-Exploitation},
  journal      = {CoRR},
  volume       = {abs/2510.18633},
  year         = {2025},
  url          = {https://doi.org/10.48550/arXiv.2510.18633},
  doi          = {10.48550/ARXIV.2510.18633},
  eprinttype    = {arXiv},
  eprint       = {2510.18633},
  bibsource    = {dblp computer science bibliography, https://dblp.org}
}

@inproceedings{zhu-etal-2025-hint,
    title = "Hint-Augmented Re-ranking: Efficient Product Search using {LLM}-Based Query Decomposition",
    author = "Zhu, Yilun  and
      Vedula, Nikhita  and
      Malmasi, Shervin",
    editor = "Inui, Kentaro  and
      Sakti, Sakriani  and
      Wang, Haofen  and
      Wong, Derek F.  and
      Bhattacharyya, Pushpak  and
      Banerjee, Biplab  and
      Ekbal, Asif  and
      Chakraborty, Tanmoy  and
      Singh, Dhirendra Pratap",
    booktitle = "Proceedings of the 14th International Joint Conference on Natural Language Processing and the 4th Conference of the Asia-Pacific Chapter of the Association for Computational Linguistics",
    month = dec,
    year = "2025",
    address = "Mumbai, India",
    publisher = "The Asian Federation of Natural Language Processing and The Association for Computational Linguistics",
    url = "https://aclanthology.org/2025.ijcnlp-short.19/",
    pages = "200--216",
    ISBN = "979-8-89176-299-2"
}

@article{DBLP:journals/corr/abs-2509-06544,
  author       = {Yunfei Zhong and
                  Jun Yang and
                  Yixing Fan and
                  Jiafeng Guo and
                  Lixin Su and
                  Maarten de Rijke and
                  Ruqing Zhang and
                  Dawei Yin and
                  Xueqi Cheng},
  title        = {Reasoning-enhanced Query Understanding through Decomposition and Interpretation},
  journal      = {CoRR},
  volume       = {abs/2509.06544},
  year         = {2025},
  url          = {https://doi.org/10.48550/arXiv.2509.06544},
  doi          = {10.48550/ARXIV.2509.06544},
  eprinttype    = {arXiv},
  eprint       = {2509.06544},
  bibsource    = {dblp computer science bibliography, https://dblp.org}
}

@inproceedings{DBLP:conf/sigir/KhattabZ20,
  author       = {Omar Khattab and
                  Matei Zaharia},
  editor       = {Jimmy X. Huang and
                  Yi Chang and
                  Xueqi Cheng and
                  Jaap Kamps and
                  Vanessa Murdock and
                  Ji{-}Rong Wen and
                  Yiqun Liu},
  title        = {ColBERT: Efficient and Effective Passage Search via Contextualized
                  Late Interaction over {BERT}},
  booktitle    = {Proceedings of the 43rd International {ACM} {SIGIR} conference on
                  research and development in Information Retrieval, {SIGIR} 2020, Virtual
                  Event, China, July 25-30, 2020},
  pages        = {39--48},
  publisher    = {{ACM}},
  year         = {2020},
  url          = {https://doi.org/10.1145/3397271.3401075},
  doi          = {10.1145/3397271.3401075},
  bibsource    = {dblp computer science bibliography, https://dblp.org}
}

@inproceedings{DBLP:conf/sigir/XiaoLZMLN24,
  author       = {Shitao Xiao and
                  Zheng Liu and
                  Peitian Zhang and
                  Niklas Muennighoff and
                  Defu Lian and
                  Jian{-}Yun Nie},
  editor       = {Grace Hui Yang and
                  Hongning Wang and
                  Sam Han and
                  Claudia Hauff and
                  Guido Zuccon and
                  Yi Zhang},
  title        = {C-Pack: Packed Resources For General Chinese Embeddings},
  booktitle    = {Proceedings of the 47th International {ACM} {SIGIR} Conference on
                  Research and Development in Information Retrieval, {SIGIR} 2024, Washington
                  DC, USA, July 14-18, 2024},
  pages        = {641--649},
  publisher    = {{ACM}},
  year         = {2024},
  url          = {https://doi.org/10.1145/3626772.3657878},
  doi          = {10.1145/3626772.3657878},
  bibsource    = {dblp computer science bibliography, https://dblp.org}
}

@inproceedings{DBLP:conf/nips/VaswaniSPUJGKP17,
  author       = {Ashish Vaswani and
                  Noam Shazeer and
                  Niki Parmar and
                  Jakob Uszkoreit and
                  Llion Jones and
                  Aidan N. Gomez and
                  Lukasz Kaiser and
                  Illia Polosukhin},
  editor       = {Isabelle Guyon and
                  Ulrike von Luxburg and
                  Samy Bengio and
                  Hanna M. Wallach and
                  Rob Fergus and
                  S. V. N. Vishwanathan and
                  Roman Garnett},
  title        = {Attention is All you Need},
  booktitle    = {Advances in Neural Information Processing Systems 30: Annual Conference
                  on Neural Information Processing Systems 2017, December 4-9, 2017,
                  Long Beach, CA, {USA}},
  pages        = {5998--6008},
  year         = {2017},
  url          = {https://proceedings.neurips.cc/paper/2017/hash/3f5ee243547dee91fbd053c1c4a845aa-Abstract.html},
  bibsource    = {dblp computer science bibliography, https://dblp.org}
}

@article{DBLP:journals/corr/abs-2506-05176,
  author       = {Yanzhao Zhang and
                  Mingxin Li and
                  Dingkun Long and
                  Xin Zhang and
                  Huan Lin and
                  Baosong Yang and
                  Pengjun Xie and
                  An Yang and
                  Dayiheng Liu and
                  Junyang Lin and
                  Fei Huang and
                  Jingren Zhou},
  title        = {Qwen3 Embedding: Advancing Text Embedding and Reranking Through Foundation
                  Models},
  journal      = {CoRR},
  volume       = {abs/2506.05176},
  year         = {2025},
  url          = {https://doi.org/10.48550/arXiv.2506.05176},
  doi          = {10.48550/ARXIV.2506.05176},
  eprinttype    = {arXiv},
  eprint       = {2506.05176},
  bibsource    = {dblp computer science bibliography, https://dblp.org}
}

@inproceedings{DBLP:conf/emnlp/NogueiraC17,
  author       = {Rodrigo Nogueira and
                  Kyunghyun Cho},
  editor       = {Martha Palmer and
                  Rebecca Hwa and
                  Sebastian Riedel},
  title        = {Task-Oriented Query Reformulation with Reinforcement Learning},
  booktitle    = {Proceedings of the 2017 Conference on Empirical Methods in Natural
                  Language Processing, {EMNLP} 2017, Copenhagen, Denmark, September
                  9-11, 2017},
  pages        = {574--583},
  publisher    = {Association for Computational Linguistics},
  year         = {2017},
  url          = {https://doi.org/10.18653/v1/d17-1061},
  doi          = {10.18653/V1/D17-1061},
  bibsource    = {dblp computer science bibliography, https://dblp.org}
}

@article{DBLP:journals/tacl/TrivediBKS22,
  author       = {Harsh Trivedi and
                  Niranjan Balasubramanian and
                  Tushar Khot and
                  Ashish Sabharwal},
  title        = {MuSiQue: Multihop Questions via Single-hop Question
                  Composition},
  journal      = {Trans. Assoc. Comput. Linguistics},
  volume       = {10},
  pages        = {539--554},
  year         = {2022},
  url          = {https://doi.org/10.1162/tacl\_a\_00475},
  doi          = {10.1162/TACL\_A\_00475},
  bibsource    = {dblp computer science bibliography, https://dblp.org}
}

@article{liu2024lost,
  title={Lost in the middle: How language models use long contexts},
  author={Liu, Nelson F. and Lin, Kevin and Hewitt, John and Paranjape, Ashwin and Bevilacqua, Michele and Petroni, Fabio and Liang, Percy},
  journal={Transactions of the Association for Computational Linguistics},
  volume={12},
  pages={157--173},
  year={2024},
  publisher={MIT Press},
  doi={10.1162/tacl_a_00638},
  url={https://doi.org/10.1162/tacl_a_00638}
}

@article{DBLP:journals/corr/abs-2601-03267,
  author       = {OpenAI},
  title        = {OpenAI {GPT-5} System Card},
  journal      = {CoRR},
  volume       = {abs/2601.03267},
  year         = {2026},
  url          = {https://doi.org/10.48550/arXiv.2601.03267},
  doi          = {10.48550/ARXIV.2601.03267},
  eprinttype   = {arXiv},
  eprint       = {2601.03267},
  bibsource    = {dblp computer science bibliography, https://dblp.org}
}

@inproceedings{lu2026beyond,
  title={Beyond Global Similarity: Multi-Conditional Retrieval for Fine-Grained Cross-Modal Understanding},
  author={Lu, Xuan and Li, Kangle and Huang, Haohang and Meng, Rui and Zeng, Wenjun and Shen, Xiaoyu},
  booktitle={Proceedings of the IEEE/CVF Conference on Computer Vision and Pattern Recognition},
  pages={9699--9709},
  year={2026}
}

@inproceedings{
lu2026rethinking,
title={Rethinking Reasoning in Document Ranking: Why Chain-of-Thought Falls Short},
author={Xuan Lu and Haohang Huang and Rui Meng and Yaohui Jin and Wenjun Zeng and Xiaoyu Shen},
booktitle={The Fourteenth International Conference on Learning Representations},
year={2026},
url={https://openreview.net/forum?id=txmqENuRcc}
}

@inproceedings{
lu2025toolsunderdocumentedsimpledocument,
title={Tools are under-documented: Simple Document Expansion Boosts Tool Retrieval},
author={Xuan Lu and Haohang Huang and Rui Meng and Yaohui Jin and Wenjun Zeng and Xiaoyu Shen},
booktitle={The Fourteenth International Conference on Learning Representations},
year={2026},
url={https://openreview.net/forum?id=g9D9MgG7iW}
}

@inproceedings{
huang2026mmeb,
title={{MMEB}-V3: Measuring the Performance Gaps of Omni-Modality Embedding Models},
author={Haohang Huang and Xuan Lu and Mingyi Su and Xuan Zhang and Ziyan Jiang and Ping Nie and Kai Zou and Tomas Pfister and Wenhu Chen and Wei Zhang and Xiaoyu Shen and Rui Meng},
booktitle={Third Conference on Language Modeling},
year={2026},
url={https://openreview.net/forum?id=EgNjgS4yOh}
}

@misc{han2026makesgoodinstructiontuningdata,
      title={What Makes Good Instruction-Tuning Data? An In-Context Learning Perspective}, 
      author={Guangzeng Han and Xiaolei Huang},
      year={2026},
      eprint={2604.25132},
      archivePrefix={arXiv},
      primaryClass={cs.CL},
      url={https://arxiv.org/abs/2604.25132}, 
}

@misc{lu2026comprankefficientllmreranking,
      title={CompRank: Efficient LLM Reranking via Token-Level Compression and Decoding-Free Scoring}, 
      author={Xuan Lu and Haohang Huang and Yingqi Fan and Junlong Tong and Yuxuan Zhang and Ping Nie and Rui Meng and Xiaoyu Shen},
      year={2026},
      eprint={2606.11700},
      archivePrefix={arXiv},
      primaryClass={cs.IR},
      url={https://arxiv.org/abs/2606.11700}, 
}

@misc{fan2026minirerankerefficientmultimodalreranking,
      title={miniReranker: Efficient Multimodal Reranking through Visual Cache Reuse and Interaction Sparsity}, 
      author={Yingqi Fan and Xuan Lu and Anhao Zhao and Junlong Tong and Ping Nie and Kai Zou and Yunpu Ma and Wei Zhang and Xiaoyu Shen},
      year={2026},
      eprint={2606.10759},
      archivePrefix={arXiv},
      primaryClass={cs.IR},
      url={https://arxiv.org/abs/2606.10759}, 
}

@article{fan2026visual,
  title={What do visual tokens really encode? uncovering sparsity and redundancy in multimodal large language models},
  author={Fan, Yingqi and Tong, Junlong and Zhao, Anhao and Shen, Xiaoyu},
  journal={arXiv preprint arXiv:2603.00510},
  year={2026}
}

@inproceedings{fan2025visipruner,
  title={Visipruner: Decoding discontinuous cross-modal dynamics for efficient multimodal llms},
  author={Fan, Yingqi and Zhao, Anhao and Fu, Jinlan and Tong, Junlong and Su, Hui and Pan, Yijie and Zhang, Wei and Shen, Xiaoyu},
  booktitle={Proceedings of the 2025 Conference on Empirical Methods in Natural Language Processing},
  pages={18896--18913},
  year={2025}
}

\onecolumn
\appendix

\section{Prompt Template for Qwen3-8B to Conduct Adaptive Decomposition}
\label{app:prompt-qwen}
In this appendix, we detail the exact prompt instructions provided to Qwen3-8B to perform adaptive query decomposition on the MultiConIR benchmark. Unlike fixed-size decomposition, the adaptive strategy allows the model to dynamically determine both the number of sub-queries and the number of conditions assigned to each sub-query, based on the semantic structure of the original query. The conversation follows the standard chat template with three roles: \texttt{system}, \texttt{user}, and \texttt{assistant}.

\vspace{0.3cm}

\begin{tcolorbox}[colback=blue!3, colframe=blue!50!black, title=\textbf{System Message}, arc=2mm, boxrule=0.5pt, left=3mm, right=3mm, top=2mm, bottom=2mm]
\small
You are an expert in complex, multi-condition information retrieval. Your task is to adaptively decompose a given complex search query into distinct, independent sub-queries to optimize downstream retrieval performance. You should dynamically determine both the number of sub-queries and how many conditions each sub-query contains, based on the semantic coherence of the original query.

\vspace{2mm}
\textbf{Rules:}
\begin{enumerate}
    \item \textbf{Adaptive Granularity:} Decide the optimal number of sub-queries based on the semantic structure of the original query.
    \item \textbf{Semantic Coherence:} Group semantically related conditions together within the same sub-query. Different sub-queries may contain different numbers of conditions, depending on how the conditions naturally cluster.
    \item \textbf{No Information Overlap:} The conditions must be mutually exclusive across sub-queries. Do not duplicate specific constraints or information across the decomposed segments.
    \item \textbf{Preserve Original Wording:} Extract and segment the text while strictly preserving the original phrasing, vocabulary, and sentence structure as much as possible. Avoid aggressively paraphrasing, rewriting, or hallucinating new information.
\end{enumerate}

\vspace{2mm}
\textbf{Output Format:} Strictly output the decomposed sub-queries in the following format, with one sub-query per line and no additional conversational text, reasoning, or thinking process: \\
\texttt{[Original\_ID]\_subq\_[Index], "[Sub-query text]"}

\vspace{2mm}
\end{tcolorbox}

\vspace{0.2cm}

\begin{tcolorbox}[colback=green!3, colframe=green!40!black, title=\textbf{User Message (Few-shot Example Input)}, arc=2mm, boxrule=0.5pt, left=3mm, right=3mm, top=2mm, bottom=2mm]
\small
\texttt{4\_Query\_8, "Find a movie matching these conditions: 1. Chaplin, dressed as tramp, sits with wife. 2. Charlie holding onions, investigates odd smell. 3. Origin: American. 4. Charlie drunkenly sees dummy as opponent. 5. Mabel reveals dummy to fighting Charlie. 6. Director: Charlie Chaplin. 7. Cast includes Charlie Chaplin, Mabel Normand. 8. Man with tennis racquet approaches wife in bar."}
\end{tcolorbox}

\vspace{0.2cm}

\begin{tcolorbox}[colback=orange!3, colframe=orange!60!black, title=\textbf{Assistant Message (Few-shot Example Output)}, arc=2mm, boxrule=0.5pt, left=3mm, right=3mm, top=2mm, bottom=2mm]
\small
\texttt{4\_Query\_8\_subq\_1, "Find a movie matching these conditions: Chaplin, dressed as tramp, sits with wife. Man with tennis racquet approaches wife in bar."} \\
\texttt{4\_Query\_8\_subq\_2, "Find a movie matching these conditions: Charlie holding onions, investigates odd smell. Charlie drunkenly sees dummy as opponent. Mabel reveals dummy to fighting Charlie."} \\
\texttt{4\_Query\_8\_subq\_3, "Find a movie matching these conditions: Origin: American. Director: Charlie Chaplin. Cast includes Charlie Chaplin, Mabel Normand."}
\end{tcolorbox}

\vspace{0.2cm}

\begin{tcolorbox}[colback=green!3, colframe=green!40!black, title=\textbf{User Message (Actual Query Input)}, arc=2mm, boxrule=0.5pt, left=3mm, right=3mm, top=2mm, bottom=2mm]
\small
\texttt{\{query\_id\}, "\{query\_text\}"}
\end{tcolorbox}

\newpage

\section{Detailed Results}
\label{sec:appendix}

\subsection{Detailed Retrieval Results}

Table~\ref{tab:main-result-retrieval} reports the average retrieval performance across the five datasets. 
For completeness, we provide the detailed results for each dataset (Books, Movies, People, Medical Cases, and Legal Documents) in the Appendix.

Tables~\ref{tab:appendix-bge-results}, ~\ref{tab:appendix-qwen0.6b-results}, ~\ref{tab:appendix-qwen4b-results} and ~\ref{tab:appendix-qwen8b-results} present the detailed results for each dataset in MultiConIR. 

\begin{table*}[h]
\centering
\small
\setlength{\tabcolsep}{4pt}
\renewcommand{\arraystretch}{1.15}
\resizebox{\textwidth}{!}{
\begin{tabular}{llcccccccccc}
\toprule
Dataset & Setting & Q1 & Q2 & Q3 & Q4 & Q5 & Q6 & Q7 & Q8 & Q9 & Q10 \\
\midrule

\multicolumn{12}{c}{\textbf{NDCG@10}} \\
\midrule

\multirow{3}{*}{\textbf{Books}}
& Baseline & 75.3 & 76.1 & 77.8 & 75.8 & 74.5 & 69.9 & 66.1 & 58.1 & 49.4 & 36.7 \\
& Decomp-RRF & 75.3 & 76.1 & 77.8 & 63.4 & 64.8 & 61.3 & 56.0 & 50.9 & 42.7 & 29.8 \\
& Decomp-Sum & 75.3 & 76.1 & 77.8 & 70.1 & 69.4 & 66.3 & 58.6 & 53.7 & 45.7 & 31.8 \\

\multirow{3}{*}{\textbf{Movies}}
& Baseline & 55.0 & 75.3 & 88.0 & 88.5 & 87.1 & 81.8 & 76.9 & 69.8 & 59.5 & 45.5 \\
& Decomp-RRF & 55.0 & 75.3 & 88.0 & 74.2 & 77.9 & 77.9 & 64.8 & 59.2 & 52.7 & 37.9 \\
& Decomp-Sum & 55.0 & 75.3 & 88.0 & 79.6 & 80.7 & 80.3 & 67.9 & 62.6 & 55.6 & 39.4 \\

\multirow{3}{*}{\textbf{People}}
& Baseline & 80.7 & 87.8 & 86.9 & 85.8 & 81.5 & 77.2 & 71.5 & 64.2 & 55.1 & 43.9 \\
& Decomp-RRF & 80.7 & 87.8 & 86.9 & 80.9 & 78.7 & 75.0 & 67.3 & 62.1 & 52.9 & 40.9 \\
& Decomp-Sum & 80.7 & 87.8 & 86.9 & 84.8 & 80.4 & 77.0 & 69.7 & 64.0 & 55.0 & 42.2 \\

\multirow{3}{*}{\textbf{Medical Cases}}
& Baseline & 61.6 & 78.6 & 83.7 & 84.3 & 81.4 & 77.9 & 72.3 & 66.9 & 59.6 & 43.8 \\
& Decomp-RRF & 61.6 & 78.6 & 83.7 & 71.2 & 72.0 & 71.5 & 63.5 & 60.0 & 51.9 & 38.9 \\
& Decomp-Sum & 61.6 & 78.6 & 83.7 & 77.8 & 76.7 & 75.1 & 67.4 & 63.1 & 55.1 & 39.4 \\

\multirow{3}{*}{\textbf{Legal Documents}}
& Baseline & 76.2 & 85.6 & 87.2 & 86.7 & 83.2 & 78.9 & 73.3 & 65.2 & 55.6 & 41.4 \\
& Decomp-RRF & 76.2 & 85.6 & 87.2 & 73.2 & 76.2 & 73.1 & 62.6 & 57.5 & 48.7 & 35.7 \\
& Decomp-Sum & 76.2 & 85.6 & 87.2 & 77.7 & 80.3 & 75.7 & 65.6 & 60.4 & 51.8 & 37.1 \\

\midrule
\multicolumn{12}{c}{\textbf{Recall@50}} \\
\midrule

\multirow{3}{*}{\textbf{Books}}
& Baseline & 89.6 & 93.0 & 95.6 & 95.4 & 97.8 & 97.9 & 98.6 & 97.9 & 98.1 & 98.9 \\
& Decomp-RRF & 89.6 & 93.0 & 95.6 & 92.8 & 95.4 & 96.1 & 94.6 & 94.5 & 94.9 & 92.1 \\
& Decomp-Sum & 89.6 & 93.0 & 95.6 & 92.8 & 95.7 & 96.3 & 95.1 & 94.6 & 95.1 & 91.7 \\

\multirow{3}{*}{\textbf{Movies}}
& Baseline & 69.2 & 87.9 & 96.8 & 99.0 & 99.8 & 99.8 & 99.9 & 99.8 & 100.0 & 100.0 \\
& Decomp-RRF & 69.2 & 87.9 & 96.8 & 91.6 & 94.9 & 98.8 & 94.3 & 96.5 & 98.4 & 98.6 \\
& Decomp-Sum & 69.2 & 87.9 & 96.8 & 91.7 & 95.0 & 98.9 & 91.8 & 94.3 & 97.8 & 94.5 \\

\multirow{3}{*}{\textbf{People}}
& Baseline & 94.5 & 98.8 & 99.4 & 99.8 & 100.0 & 100.0 & 100.0 & 100.0 & 99.9 & 100.0 \\
& Decomp-RRF & 94.5 & 98.8 & 99.4 & 99.8 & 100.0 & 99.9 & 99.8 & 100.0 & 100.0 & 99.5 \\
& Decomp-Sum & 94.5 & 98.8 & 99.4 & 100.0 & 100.0 & 100.0 & 99.8 & 100.0 & 100.0 & 99.8 \\

\multirow{3}{*}{\textbf{Medical Cases}}
& Baseline & 82.7 & 95.0 & 99.0 & 99.7 & 100.0 & 100.0 & 100.0 & 100.0 & 100.0 & 100.0 \\
& Decomp-RRF & 82.7 & 95.0 & 99.0 & 97.2 & 97.8 & 98.7 & 97.6 & 98.5 & 98.1 & 97.6 \\
& Decomp-Sum & 82.7 & 95.0 & 99.0 & 97.4 & 97.9 & 98.7 & 97.1 & 98.6 & 97.6 & 93.7 \\

\multirow{3}{*}{\textbf{Legal Documents}}
& Baseline & 92.9 & 97.7 & 97.6 & 99.5 & 99.8 & 99.6 & 100.0 & 99.8 & 100.0 & 100.0 \\
& Decomp-RRF & 92.9 & 97.7 & 97.6 & 94.4 & 98.4 & 98.0 & 93.5 & 96.9 & 97.6 & 92.7 \\
& Decomp-Sum & 92.9 & 97.7 & 97.6 & 94.6 & 98.1 & 98.2 & 93.8 & 96.8 & 97.8 & 90.7 \\

\bottomrule
\end{tabular}
}
\caption{Detailed retrieval results for each dataset using \textbf{BGE-large-en-v1.5}. Values are scaled by 100.}
\label{tab:appendix-bge-results}
\end{table*}

\begin{table*}[h]
\centering
\small
\setlength{\tabcolsep}{4pt}
\renewcommand{\arraystretch}{1.15}
\resizebox{\textwidth}{!}{
\begin{tabular}{llcccccccccc}
\toprule
Dataset & Setting & Q1 & Q2 & Q3 & Q4 & Q5 & Q6 & Q7 & Q8 & Q9 & Q10 \\
\midrule

\multicolumn{12}{c}{\textbf{NDCG@10}} \\
\midrule

\multirow{3}{*}{\textbf{Books}}
& Baseline & 67.9 & 82.7 & 82.4 & 82.4 & 79.0 & 77.1 & 70.8 & 64.0 & 55.6 & 43.5 \\
& Decomp-RRF & 67.9 & 82.7 & 82.4 & 73.6 & 74.1 & 71.7 & 65.4 & 58.3 & 52.1 & 37.8 \\
& Decomp-Sum & 67.9 & 82.7 & 82.4 & 78.4 & 76.3 & 75.0 & 68.5 & 62.3 & 54.7 & 39.2 \\

\multirow{3}{*}{\textbf{Movies}}
& Baseline & 62.1 & 84.1 & 89.3 & 88.4 & 85.3 & 81.0 & 75.8 & 68.4 & 59.0 & 45.1 \\
& Decomp-RRF & 62.1 & 84.1 & 89.3 & 78.6 & 79.9 & 75.8 & 67.2 & 62.3 & 54.1 & 39.1 \\
& Decomp-Sum & 62.1 & 84.1 & 89.3 & 82.5 & 83.0 & 79.1 & 70.0 & 66.0 & 56.7 & 42.9 \\

\multirow{3}{*}{\textbf{People}}
& Baseline & 78.2 & 90.6 & 91.3 & 88.5 & 84.5 & 80.3 & 75.7 & 68.4 & 61.1 & 50.6 \\
& Decomp-RRF & 78.2 & 90.6 & 91.3 & 83.7 & 80.5 & 77.6 & 71.2 & 64.8 & 58.1 & 45.5 \\
& Decomp-Sum & 78.2 & 90.6 & 91.3 & 86.8 & 81.4 & 78.5 & 73.5 & 66.5 & 59.0 & 47.0 \\

\multirow{3}{*}{\textbf{Medical Cases}}
& Baseline & 57.9 & 77.8 & 82.1 & 83.2 & 80.5 & 76.9 & 71.7 & 66.6 & 59.3 & 46.0 \\
& Decomp-RRF & 57.9 & 77.8 & 82.1 & 77.8 & 76.3 & 73.3 & 67.4 & 62.5 & 54.1 & 39.8 \\
& Decomp-Sum & 57.9 & 77.8 & 82.1 & 81.2 & 79.2 & 75.6 & 70.1 & 65.2 & 57.0 & 41.4 \\

\multirow{3}{*}{\textbf{Legal Documents}}
& Baseline & 72.5 & 84.9 & 85.8 & 86.5 & 83.0 & 78.1 & 73.2 & 65.1 & 57.6 & 42.3 \\
& Decomp-RRF & 72.5 & 84.9 & 85.8 & 77.5 & 78.2 & 75.7 & 65.9 & 62.6 & 54.8 & 39.1 \\
& Decomp-Sum & 72.5 & 84.9 & 85.8 & 83.3 & 81.6 & 77.5 & 70.8 & 65.7 & 56.5 & 42.0 \\

\midrule
\multicolumn{12}{c}{\textbf{Recall@50}} \\
\midrule

\multirow{3}{*}{\textbf{Books}}
& Baseline & 86.6 & 96.4 & 97.3 & 98.9 & 99.6 & 100.0 & 100.0 & 99.9 & 99.8 & 99.8 \\
& Decomp-RRF & 86.6 & 96.4 & 97.3 & 97.2 & 97.5 & 99.3 & 97.9 & 98.9 & 99.4 & 97.5 \\
& Decomp-Sum & 86.6 & 96.4 & 97.3 & 96.8 & 97.5 & 99.5 & 98.4 & 99.2 & 99.6 & 97.9 \\

\multirow{3}{*}{\textbf{Movies}}
& Baseline & 75.4 & 93.3 & 99.0 & 99.4 & 99.7 & 99.5 & 100.0 & 100.0 & 100.0 & 100.0 \\
& Decomp-RRF & 75.4 & 93.3 & 99.0 & 94.6 & 98.1 & 97.8 & 96.5 & 98.4 & 99.3 & 99.0 \\
& Decomp-Sum & 75.4 & 93.3 & 99.0 & 94.5 & 98.1 & 97.7 & 96.5 & 98.7 & 99.4 & 98.2 \\

\multirow{3}{*}{\textbf{People}}
& Baseline & 93.6 & 100.0 & 100.0 & 100.0 & 100.0 & 100.0 & 100.0 & 100.0 & 100.0 & 100.0 \\
& Decomp-RRF & 93.6 & 100.0 & 100.0 & 99.9 & 100.0 & 100.0 & 100.0 & 100.0 & 100.0 & 100.0 \\
& Decomp-Sum & 93.6 & 100.0 & 100.0 & 99.9 & 100.0 & 100.0 & 100.0 & 100.0 & 100.0 & 100.0 \\

\multirow{3}{*}{\textbf{Medical Cases}}
& Baseline & 83.5 & 96.4 & 98.8 & 99.6 & 99.8 & 100.0 & 100.0 & 100.0 & 100.0 & 100.0 \\
& Decomp-RRF & 83.5 & 96.4 & 98.8 & 98.7 & 99.2 & 99.6 & 99.4 & 99.6 & 99.5 & 99.3 \\
& Decomp-Sum & 83.5 & 96.4 & 98.8 & 98.9 & 99.2 & 99.7 & 99.5 & 99.6 & 99.5 & 99.8 \\

\multirow{3}{*}{\textbf{Legal Documents}}
& Baseline & 89.5 & 96.5 & 98.0 & 99.3 & 99.4 & 100.0 & 100.0 & 100.0 & 100.0 & 100.0 \\
& Decomp-RRF & 89.5 & 96.5 & 98.0 & 97.0 & 98.5 & 99.2 & 98.1 & 99.8 & 98.8 & 98.2 \\
& Decomp-Sum & 89.5 & 96.5 & 98.0 & 97.2 & 98.5 & 99.3 & 99.3 & 99.8 & 99.5 & 99.0 \\

\bottomrule
\end{tabular}
}
\caption{Detailed retrieval results for each dataset using \textbf{Qwen3-Embedding-0.6B}. Values are scaled by 100.}
\label{tab:appendix-qwen0.6b-results}
\end{table*}

\begin{table*}[h]
\centering
\small
\setlength{\tabcolsep}{4pt}
\renewcommand{\arraystretch}{1.15}
\resizebox{\textwidth}{!}{
\begin{tabular}{llcccccccccc}
\toprule
Dataset & Setting & Q1 & Q2 & Q3 & Q4 & Q5 & Q6 & Q7 & Q8 & Q9 & Q10 \\
\midrule

\multicolumn{12}{c}{\textbf{NDCG@10}} \\
\midrule

\multirow{3}{*}{\textbf{Books}}
& Baseline & 74.1 & 85.1 & 86.5 & 85.2 & 82.3 & 77.8 & 72.1 & 65.1 & 55.9 & 43.2 \\
& Decomp-RRF & 74.1 & 85.1 & 86.5 & 78.8 & 77.6 & 73.7 & 66.5 & 59.9 & 51.3 & 39.1 \\
& Decomp-Sum & 74.1 & 85.1 & 86.5 & 83.3 & 80.6 & 76.0 & 69.1 & 62.7 & 53.7 & 40.4 \\

\multirow{3}{*}{\textbf{Movies}}
& Baseline & 53.1 & 76.4 & 85.7 & 86.3 & 84.7 & 80.0 & 74.8 & 66.8 & 56.3 & 42.9 \\
& Decomp-RRF & 53.1 & 76.4 & 85.7 & 71.8 & 74.6 & 75.5 & 63.3 & 60.4 & 54.0 & 39.4 \\
& Decomp-Sum & 53.1 & 76.4 & 85.7 & 78.1 & 79.5 & 78.6 & 68.9 & 64.3 & 58.3 & 43.8 \\

\multirow{3}{*}{\textbf{People}}
& Baseline & 82.8 & 91.6 & 91.1 & 89.0 & 84.3 & 80.2 & 73.7 & 66.9 & 58.8 & 44.2 \\
& Decomp-RRF & 82.8 & 91.6 & 91.1 & 84.8 & 81.7 & 76.9 & 68.4 & 61.8 & 54.0 & 39.7 \\
& Decomp-Sum & 82.8 & 91.6 & 91.1 & 87.3 & 83.1 & 78.3 & 71.1 & 63.5 & 55.7 & 41.4 \\

\multirow{3}{*}{\textbf{Medical Cases}}
& Baseline & 62.5 & 79.0 & 83.5 & 83.0 & 78.4 & 74.7 & 69.7 & 65.0 & 56.5 & 44.8 \\
& Decomp-RRF & 62.5 & 79.0 & 83.5 & 79.4 & 76.0 & 72.5 & 65.2 & 61.0 & 52.1 & 41.9 \\
& Decomp-Sum & 62.5 & 79.0 & 83.5 & 83.0 & 77.7 & 73.7 & 67.5 & 62.8 & 53.7 & 44.1 \\

\multirow{3}{*}{\textbf{Legal Documents}}
& Baseline & 76.7 & 86.5 & 87.8 & 86.5 & 83.0 & 78.0 & 73.3 & 65.7 & 56.0 & 39.6 \\
& Decomp-RRF & 76.7 & 86.5 & 87.8 & 81.1 & 80.8 & 76.2 & 70.4 & 64.2 & 55.4 & 41.9 \\
& Decomp-Sum & 76.7 & 86.5 & 87.8 & 85.1 & 82.5 & 77.9 & 73.6 & 67.7 & 57.5 & 45.7 \\

\midrule
\multicolumn{12}{c}{\textbf{Recall@50}} \\
\midrule

\multirow{3}{*}{\textbf{Books}}
& Baseline & 89.9 & 97.4 & 99.0 & 99.5 & 99.9 & 99.9 & 100.0 & 100.0 & 99.8 & 100.0 \\
& Decomp-RRF & 89.9 & 97.4 & 99.0 & 98.7 & 99.3 & 99.7 & 99.3 & 99.8 & 99.3 & 99.6 \\
& Decomp-Sum & 89.9 & 97.4 & 99.0 & 98.7 & 99.3 & 99.7 & 99.6 & 99.8 & 99.5 & 99.6 \\

\multirow{3}{*}{\textbf{Movies}}
& Baseline & 64.5 & 86.4 & 96.2 & 98.1 & 99.7 & 99.5 & 100.0 & 99.9 & 100.0 & 100.0 \\
& Decomp-RRF & 64.5 & 86.4 & 96.2 & 89.5 & 93.6 & 97.7 & 91.1 & 95.0 & 96.7 & 98.8 \\
& Decomp-Sum & 64.5 & 86.4 & 96.2 & 89.5 & 93.7 & 97.7 & 94.2 & 97.5 & 98.5 & 98.2 \\

\multirow{3}{*}{\textbf{People}}
& Baseline & 95.9 & 99.8 & 100.0 & 100.0 & 100.0 & 100.0 & 100.0 & 100.0 & 100.0 & 100.0 \\
& Decomp-RRF & 95.9 & 99.8 & 100.0 & 100.0 & 100.0 & 100.0 & 100.0 & 100.0 & 100.0 & 100.0 \\
& Decomp-Sum & 95.9 & 99.8 & 100.0 & 100.0 & 100.0 & 100.0 & 100.0 & 100.0 & 100.0 & 100.0 \\

\multirow{3}{*}{\textbf{Medical Cases}}
& Baseline & 88.2 & 97.5 & 99.7 & 99.8 & 99.9 & 99.8 & 99.9 & 100.0 & 100.0 & 100.0 \\
& Decomp-RRF & 88.2 & 97.5 & 99.7 & 99.6 & 100.0 & 99.5 & 99.1 & 99.8 & 99.5 & 99.8 \\
& Decomp-Sum & 88.2 & 97.5 & 99.7 & 99.7 & 100.0 & 99.6 & 99.6 & 99.8 & 99.5 & 99.8 \\

\multirow{3}{*}{\textbf{Legal Documents}}
& Baseline & 92.2 & 97.4 & 98.4 & 99.3 & 100.0 & 100.0 & 100.0 & 99.9 & 100.0 & 99.8 \\
& Decomp-RRF & 92.2 & 97.4 & 98.4 & 98.4 & 99.3 & 99.6 & 99.4 & 99.8 & 99.6 & 99.0 \\
& Decomp-Sum & 92.2 & 97.4 & 98.4 & 98.5 & 99.3 & 99.5 & 99.6 & 99.8 & 99.8 & 99.2 \\

\bottomrule
\end{tabular}
}
\caption{Detailed retrieval results for each dataset using \textbf{Qwen3-Embedding-4B}. Values are scaled by 100.}
\label{tab:appendix-qwen4b-results}
\end{table*}

\begin{table*}[h]
\centering
\small
\setlength{\tabcolsep}{4pt}
\renewcommand{\arraystretch}{1.15}
\resizebox{\textwidth}{!}{
\begin{tabular}{llcccccccccc}
\toprule
Dataset & Setting & Q1 & Q2 & Q3 & Q4 & Q5 & Q6 & Q7 & Q8 & Q9 & Q10 \\
\midrule

\multicolumn{12}{c}{\textbf{NDCG@10}} \\
\midrule

\multirow{3}{*}{\textbf{Books}} & Baseline & 76.0 & 85.3 & 85.9 & 84.1 & 82.0 & 77.7 & 71.5 & 65.2 & 56.0 & 47.1 \\
& Decomp-RRF & 76.0 & 85.3 & 85.9 & 79.4 & 78.4 & 74.8 & 67.8 & 62.0 & 53.7 & 45.1 \\
& Decomp-Sum & 76.0 & 85.3 & 85.9 & 82.8 & 80.3 & 76.5 & 69.7 & 63.1 & 55.0 & 46.9 \\

\multirow{3}{*}{\textbf{Movies}} & Baseline & 57.8 & 79.8 & 88.8 & 89.3 & 88.1 & 83.9 & 79.3 & 72.5 & 62.9 & 45.8 \\
& Decomp-RRF & 57.8 & 79.8 & 88.8 & 76.5 & 79.3 & 77.8 & 67.1 & 63.8 & 58.0 & 39.3 \\
& Decomp-Sum & 57.8 & 79.8 & 88.8 & 81.2 & 82.9 & 81.5 & 72.6 & 68.5 & 62.9 & 44.3 \\

\multirow{3}{*}{\textbf{People}} & Baseline & 87.4 & 93.0 & 91.3 & 89.0 & 85.8 & 81.5 & 75.4 & 69.1 & 62.2 & 51.4 \\
& Decomp-RRF & 87.4 & 93.0 & 91.3 & 84.9 & 82.2 & 77.8 & 71.5 & 64.8 & 57.6 & 43.6 \\
& Decomp-Sum & 87.4 & 93.0 & 91.3 & 87.5 & 83.2 & 79.2 & 72.7 & 65.7 & 57.9 & 44.9 \\

\multirow{3}{*}{\textbf{Medical Cases}} & Baseline & 66.2 & 82.7 & 85.0 & 84.1 & 80.0 & 75.7 & 69.6 & 64.2 & 58.0 & 45.6 \\
& Decomp-RRF & 66.2 & 82.7 & 85.0 & 78.4 & 76.1 & 72.2 & 65.6 & 60.5 & 53.7 & 44.0 \\
& Decomp-Sum & 66.2 & 82.7 & 85.0 & 82.4 & 78.3 & 74.2 & 67.8 & 63.0 & 55.9 & 46.1 \\

\multirow{3}{*}{\textbf{Legal Documents}} & Baseline & 75.4 & 84.5 & 86.1 & 85.0 & 81.6 & 77.6 & 71.9 & 63.9 & 54.2 & 37.3 \\
& Decomp-RRF & 75.4 & 84.5 & 86.1 & 79.2 & 77.7 & 74.6 & 68.3 & 59.7 & 50.6 & 36.8 \\
& Decomp-Sum & 75.4 & 84.5 & 86.1 & 83.3 & 80.2 & 76.6 & 70.9 & 61.4 & 53.4 & 39.5 \\

\midrule
\multicolumn{12}{c}{\textbf{Recall@50}} \\
\midrule

\multirow{3}{*}{\textbf{Books}} & Baseline & 91.3 & 98.0 & 99.7 & 99.8 & 100.0 & 100.0 & 100.0 & 100.0 & 99.8 & 100.0 \\
& Decomp-RRF & 91.3 & 98.0 & 99.7 & 99.0 & 99.7 & 99.9 & 99.6 & 99.9 & 99.6 & 99.8 \\
& Decomp-Sum & 91.3 & 98.0 & 99.7 & 99.1 & 99.7 & 99.9 & 99.7 & 99.9 & 99.7 & 99.8 \\

\multirow{3}{*}{\textbf{Movies}} & Baseline & 69.4 & 88.8 & 96.5 & 98.9 & 99.8 & 99.9 & 100.0 & 100.0 & 100.0 & 100.0 \\
& Decomp-RRF & 69.4 & 88.8 & 96.5 & 90.1 & 95.0 & 98.3 & 94.0 & 97.2 & 98.5 & 99.2 \\
& Decomp-Sum & 69.4 & 88.8 & 96.5 & 90.4 & 95.0 & 98.1 & 94.3 & 98.0 & 99.2 & 97.8 \\

\multirow{3}{*}{\textbf{People}} & Baseline & 97.2 & 100.0 & 100.0 & 100.0 & 100.0 & 100.0 & 100.0 & 100.0 & 100.0 & 100.0 \\
& Decomp-RRF & 97.2 & 100.0 & 100.0 & 100.0 & 100.0 & 100.0 & 100.0 & 100.0 & 100.0 & 100.0 \\
& Decomp-Sum & 97.2 & 100.0 & 100.0 & 100.0 & 100.0 & 100.0 & 100.0 & 100.0 & 100.0 & 100.0 \\

\multirow{3}{*}{\textbf{Medical Cases}} & Baseline & 89.3 & 98.7 & 99.8 & 100.0 & 100.0 & 100.0 & 100.0 & 100.0 & 100.0 & 100.0 \\
& Decomp-RRF & 89.3 & 98.7 & 99.8 & 99.4 & 100.0 & 99.9 & 99.8 & 99.8 & 99.5 & 99.8 \\
& Decomp-Sum & 89.3 & 98.7 & 99.8 & 99.6 & 100.0 & 100.0 & 99.9 & 99.9 & 99.9 & 100.0 \\

\multirow{3}{*}{\textbf{Legal Documents}} & Baseline & 91.5 & 97.6 & 98.8 & 99.3 & 99.9 & 100.0 & 100.0 & 99.8 & 100.0 & 100.0 \\
& Decomp-RRF & 91.5 & 97.6 & 98.8 & 98.5 & 99.5 & 99.6 & 99.3 & 99.0 & 99.1 & 98.5 \\
& Decomp-Sum & 91.5 & 97.6 & 98.8 & 98.7 & 99.7 & 99.8 & 100.0 & 99.3 & 100.0 & 99.5 \\

\bottomrule
\end{tabular}
}
\caption{Detailed retrieval results for each dataset using \textbf{Qwen3-Embedding-8B}. Values are scaled by 100.}
\label{tab:appendix-qwen8b-results}
\end{table*}

\clearpage %

\subsection{Detailed Reranking Results}
Table~\ref{tab:main-result-reranking} reports the average reranking performance across the five datasets. 
For completeness, we provide the detailed results for each dataset (Books, Movies, People, Medical Cases, and Legal Documents) in the Appendix.

Tables ~\ref{tab:appendix-reranking-qwen0.6b-results}, ~\ref{tab:appendix-reranking-qwen4b-results} and~\ref{tab:appendix-reranking-qwen8b-results} present the detailed results for each dataset in MultiConIR.

\begin{table*}[h]
\centering
\small
\setlength{\tabcolsep}{4pt}
\renewcommand{\arraystretch}{1.15}
\resizebox{\textwidth}{!}{
\begin{tabular}{llcccccccccc}
\toprule
Dataset & Setting & Q1 & Q2 & Q3 & Q4 & Q5 & Q6 & Q7 & Q8 & Q9 & Q10 \\
\midrule
\multicolumn{12}{c}{\textbf{NDCG@10}} \\
\midrule
\multirow{4}{*}{\textbf{Books}}
& Orig. Base    & 75.3 & 76.1 & 77.8 & 75.8 & 74.5 & 69.9 & 66.1 & 58.1 & 49.4 & 36.7 \\
& Pure Rerank   & 85.2 & 88.0 & 86.9 & 82.1 & 79.8 & 73.7 & 68.0 & 59.6 & 51.4 & 43.8 \\
& Decomp-Sum    & 85.2 & 88.0 & 86.9 & 85.6 & 84.9 & 80.0 & 77.0 & 71.6 & 64.0 & 55.9 \\
& Decomp-RRF    & 85.2 & 88.0 & 86.9 & 84.6 & 83.9 & 79.3 & 76.1 & 71.2 & 63.3 & 58.7 \\
\multirow{4}{*}{\textbf{Movies}}
& Orig. Base    & 55.0 & 75.3 & 88.0 & 88.5 & 87.1 & 81.8 & 76.9 & 69.8 & 59.5 & 45.5 \\
& Pure Rerank   & 65.9 & 83.4 & 87.3 & 82.6 & 76.8 & 69.9 & 56.8 & 49.9 & 49.9 & 38.4 \\
& Decomp-Sum    & 65.9 & 83.4 & 87.3 & 90.8 & 85.6 & 79.6 & 76.6 & 70.7 & 58.8 & 52.9 \\
& Decomp-RRF    & 65.9 & 83.4 & 87.3 & 88.8 & 84.5 & 78.0 & 74.1 & 66.7 & 56.0 & 47.1 \\
\multirow{4}{*}{\textbf{People}}
& Orig. Base    & 80.7 & 87.8 & 86.9 & 85.8 & 81.5 & 77.2 & 71.5 & 64.2 & 55.1 & 43.9 \\
& Pure Rerank   & 91.1 & 92.8 & 89.0 & 84.6 & 79.2 & 74.6 & 66.4 & 59.5 & 51.5 & 39.2 \\
& Decomp-Sum    & 91.1 & 92.8 & 89.0 & 89.5 & 85.3 & 81.9 & 78.3 & 72.1 & 64.1 & 53.3 \\
& Decomp-RRF    & 91.1 & 92.8 & 89.0 & 89.0 & 85.3 & 82.2 & 77.9 & 72.1 & 63.9 & 53.9 \\
\multirow{4}{*}{\textbf{Medical Cases}}
& Orig. Base    & 61.6 & 78.6 & 83.7 & 84.3 & 81.4 & 77.9 & 72.3 & 66.9 & 59.6 & 43.8 \\
& Pure Rerank   & 72.9 & 87.5 & 88.1 & 83.3 & 77.7 & 71.8 & 64.4 & 56.3 & 50.1 & 40.1 \\
& Decomp-Sum    & 72.9 & 87.5 & 88.1 & 90.0 & 85.2 & 80.8 & 76.2 & 70.5 & 62.1 & 53.4 \\
& Decomp-RRF    & 72.9 & 87.5 & 88.1 & 87.8 & 84.4 & 80.5 & 74.2 & 68.0 & 60.8 & 51.7 \\
\multirow{4}{*}{\textbf{Legal Documents}}
& Orig. Base    & 76.2 & 85.6 & 87.2 & 86.7 & 83.2 & 78.9 & 73.3 & 65.2 & 55.6 & 41.4 \\
& Pure Rerank   & 85.0 & 86.8 & 82.4 & 80.0 & 74.8 & 70.4 & 65.0 & 56.9 & 50.4 & 38.3 \\
& Decomp-Sum    & 85.0 & 86.8 & 82.4 & 86.8 & 81.3 & 75.1 & 72.8 & 66.5 & 58.2 & 48.4 \\
& Decomp-RRF    & 85.0 & 86.8 & 82.4 & 85.2 & 80.4 & 75.3 & 73.1 & 66.3 & 58.4 & 48.9 \\
\bottomrule
\end{tabular}
}
\caption{Detailed reranking results for each dataset using \textbf{Qwen3-Reranker-0.6B}. Values are scaled by 100.}
\label{tab:appendix-reranking-qwen0.6b-results}
\end{table*}

\begin{table*}[h]
\centering
\small
\setlength{\tabcolsep}{5pt}
\renewcommand{\arraystretch}{1.15}
\resizebox{\textwidth}{!}{
\begin{tabular}{llcccccccccc}
\toprule
Dataset & Setting & Q1 & Q2 & Q3 & Q4 & Q5 & Q6 & Q7 & Q8 & Q9 & Q10 \\
\midrule
\multicolumn{12}{c}{\textbf{NDCG@10}} \\
\midrule
\multirow{4}{*}{\textbf{Books}}
& Orig. Base    & 75.3 & 76.1 & 77.8 & 75.8 & 74.5 & 69.9 & 66.1 & 58.1 & 49.4 & 36.7 \\
& Pure Rerank   & 86.4 & 89.2 & 88.4 & 84.3 & 81.6 & 76.4 & 70.6 & 64.1 & 56.6 & 52.7 \\
& Decomp-Sum    & 86.4 & 89.2 & 88.4 & 86.6 & 84.9 & 81.1 & 75.4 & 68.5 & 62.3 & 59.0 \\
& Decomp-RRF    & 86.4 & 89.2 & 88.4 & 84.9 & 83.5 & 79.5 & 73.2 & 67.6 & 61.0 & 58.3 \\
\multirow{4}{*}{\textbf{Movies}}
& Orig. Base    & 55.0 & 75.3 & 88.0 & 88.5 & 87.1 & 81.8 & 76.9 & 69.8 & 59.5 & 45.5 \\
& Pure Rerank   & 66.3 & 85.3 & 91.0 & 88.7 & 83.2 & 77.4 & 71.2 & 63.3 & 53.9 & 42.2 \\
& Decomp-Sum    & 66.3 & 85.3 & 91.0 & 91.4 & 89.0 & 84.9 & 81.4 & 74.0 & 66.7 & 56.0 \\
& Decomp-RRF    & 66.3 & 85.3 & 91.0 & 89.7 & 87.2 & 83.3 & 78.3 & 71.3 & 62.5 & 52.8 \\
\multirow{4}{*}{\textbf{People}}
& Orig. Base    & 80.7 & 87.8 & 86.9 & 85.8 & 81.5 & 77.2 & 71.5 & 64.2 & 55.1 & 43.9 \\
& Pure Rerank   & 93.0 & 93.9 & 90.7 & 85.5 & 80.4 & 76.0 & 69.8 & 62.8 & 56.9 & 45.6 \\
& Decomp-Sum    & 93.0 & 93.9 & 90.7 & 90.5 & 84.8 & 80.5 & 77.7 & 71.6 & 65.5 & 58.9 \\
& Decomp-RRF    & 93.0 & 93.9 & 90.7 & 88.8 & 83.2 & 79.5 & 75.9 & 68.7 & 63.5 & 55.1 \\
\multirow{4}{*}{\textbf{Medical Cases}}
& Orig. Base    & 61.6 & 78.6 & 83.7 & 84.3 & 81.4 & 77.9 & 72.3 & 66.9 & 59.6 & 43.8 \\
& Pure Rerank   & 74.2 & 89.4 & 92.0 & 89.1 & 85.0 & 79.9 & 72.8 & 66.7 & 58.7 & 50.6 \\
& Decomp-Sum    & 74.2 & 89.4 & 92.0 & 91.0 & 87.3 & 82.4 & 78.0 & 74.7 & 64.4 & 58.6 \\
& Decomp-RRF    & 74.2 & 89.4 & 92.0 & 89.2 & 85.8 & 81.1 & 74.4 & 68.4 & 62.1 & 54.4 \\
\multirow{4}{*}{\textbf{Legal Documents}}
& Orig. Base    & 76.2 & 85.6 & 87.2 & 86.7 & 83.2 & 78.9 & 73.3 & 65.2 & 55.6 & 41.4 \\
& Pure Rerank   & 88.0 & 89.7 & 86.9 & 83.8 & 78.3 & 73.7 & 67.5 & 59.1 & 51.6 & 40.4 \\
& Decomp-Sum    & 88.0 & 89.7 & 86.9 & 88.6 & 84.8 & 78.7 & 75.7 & 69.4 & 62.2 & 53.2 \\
& Decomp-RRF    & 88.0 & 89.7 & 86.9 & 87.4 & 83.5 & 78.3 & 74.4 & 66.6 & 59.1 & 51.1 \\
\bottomrule
\end{tabular}
}
\caption{Detailed reranking results for each dataset using \textbf{Qwen3-Reranker-4B}. Values are scaled by 100.}
\label{tab:appendix-reranking-qwen4b-results}
\end{table*}

\begin{table*}[h]
\centering
\small
\setlength{\tabcolsep}{5pt}
\renewcommand{\arraystretch}{1.15}
\resizebox{\textwidth}{!}{
\begin{tabular}{llcccccccccc}
\toprule
Dataset & Setting & Q1 & Q2 & Q3 & Q4 & Q5 & Q6 & Q7 & Q8 & Q9 & Q10 \\
\midrule
\multicolumn{12}{c}{\textbf{NDCG@10}} \\
\midrule
\multirow{4}{*}{\textbf{Books}}
& Orig. Base    & 75.3 & 76.1 & 77.8 & 75.8 & 74.5 & 69.9 & 66.1 & 58.1 & 49.4 & 36.7 \\
& Pure Rerank   & 86.5 & 89.1 & 88.8 & 85.0 & 82.6 & 77.4 & 72.1 & 65.7 & 60.9 & 60.8 \\
& Decomp-Sum    & 86.5 & 89.1 & 88.8 & 87.4 & 84.5 & 81.1 & 76.5 & 69.9 & 65.0 & 66.5 \\
& Decomp-RRF    & 86.5 & 89.1 & 88.8 & 84.6 & 83.2 & 79.0 & 73.3 & 67.5 & 62.1 & 68.2 \\
\multirow{4}{*}{\textbf{Movies}}
& Orig. Base    & 55.0 & 75.3 & 88.0 & 88.5 & 87.1 & 81.8 & 76.9 & 69.8 & 59.5 & 45.5 \\
& Pure Rerank   & 65.5 & 86.0 & 93.1 & 93.5 & 91.0 & 87.4 & 82.9 & 75.2 & 66.0 & 53.7 \\
& Decomp-Sum    & 65.5 & 86.0 & 93.1 & 93.7 & 91.0 & 88.6 & 84.5 & 79.9 & 74.4 & 63.8 \\
& Decomp-RRF    & 65.5 & 86.0 & 93.1 & 90.2 & 88.8 & 86.2 & 80.9 & 75.9 & 70.8 & 59.9 \\
\multirow{4}{*}{\textbf{People}}
& Orig. Base    & 80.7 & 87.8 & 86.9 & 85.8 & 81.5 & 77.2 & 71.5 & 64.2 & 55.1 & 43.9 \\
& Pure Rerank   & 92.5 & 95.7 & 93.5 & 91.4 & 88.0 & 83.3 & 77.0 & 70.4 & 60.1 & 50.2 \\
& Decomp-Sum    & 92.5 & 95.7 & 93.5 & 92.1 & 88.1 & 85.7 & 81.1 & 75.9 & 72.0 & 70.9 \\
& Decomp-RRF    & 92.5 & 95.7 & 93.5 & 89.7 & 87.3 & 84.1 & 79.7 & 74.2 & 70.0 & 71.9 \\
\multirow{4}{*}{\textbf{Medical Cases}}
& Orig. Base    & 61.6 & 78.6 & 83.7 & 84.3 & 81.4 & 77.9 & 72.3 & 66.9 & 59.6 & 43.8 \\
& Pure Rerank   & 74.3 & 91.0 & 94.8 & 92.6 & 88.0 & 82.9 & 76.3 & 67.1 & 58.3 & 46.6 \\
& Decomp-Sum    & 74.3 & 91.0 & 94.8 & 94.3 & 91.0 & 88.9 & 83.9 & 79.7 & 73.5 & 69.7 \\
& Decomp-RRF    & 74.3 & 91.0 & 94.8 & 91.8 & 89.1 & 86.4 & 80.1 & 75.6 & 68.4 & 67.7 \\
\multirow{4}{*}{\textbf{Legal Documents}}
& Orig. Base    & 76.2 & 85.6 & 87.2 & 86.7 & 83.2 & 78.9 & 73.3 & 65.2 & 55.6 & 41.4 \\
& Pure Rerank   & 87.5 & 92.1 & 90.3 & 89.0 & 85.1 & 79.2 & 74.3 & 65.6 & 58.7 & 48.3 \\
& Decomp-Sum    & 87.5 & 92.1 & 90.3 & 90.4 & 87.2 & 81.7 & 80.2 & 73.2 & 65.3 & 61.9 \\
& Decomp-RRF    & 87.5 & 92.1 & 90.3 & 87.8 & 85.2 & 80.1 & 76.6 & 69.6 & 64.2 & 63.0 \\
\bottomrule
\end{tabular}
}
\caption{Detailed reranking results for each dataset using \textbf{Qwen3-Reranker-8B}. Values are scaled by 100.}
\label{tab:appendix-reranking-qwen8b-results}
\end{table*}

\clearpage %

\subsection{Detailed Recall Failure Rate In Retrieval}
Figure~\ref{fig:recall_failure} reports the average recall failure rate of sub-queries in retrieval process across the five datasets. 
For completeness, we provide the detailed results for each dataset (Books, Movies, People, Medical Cases, and Legal Documents) in the Appendix.

Tables ~\ref{tab:subq-recall-failure-transposed} and Table ~\ref{tab:baseline-recall-failure-transposed} present the detailed results for each dataset in MultiConIR. 
\begin{table*}[h]
\centering
\small
\setlength{\tabcolsep}{8pt}
\renewcommand{\arraystretch}{1.15}
\resizebox{\textwidth}{!}{
\begin{tabular}{lcccccccccc}
\toprule
\textbf{Dataset} & \textbf{Q1} & \textbf{Q2} & \textbf{Q3} & \textbf{Q4} & \textbf{Q5} & \textbf{Q6} & \textbf{Q7} & \textbf{Q8} & \textbf{Q9} & \textbf{Q10} \\
\midrule
\textbf{Books} & 7.9 & 4.2 & 2.5 & 10.3 & 7.3 & 7.7 & 11.3 & 9.7 & 11.1 & 16.1 \\
\textbf{People} & 3.6 & 0.2 & 0.2 & 1.3 & 0.6 & 0.7 & 1.7 & 1.1 & 1.3 & 1.9 \\
\textbf{Movies} & 23.0 & 8.8 & 2.2 & 10.6 & 6.0 & 3.2 & 10.6 & 7.5 & 4.7 & 11.7 \\
\textbf{Medical Cases} & 12.5 & 3.3 & 0.4 & 7.7 & 6.0 & 4.0 & 10.6 & 7.9 & 8.2 & 12.1 \\
\textbf{Legal Documents} & 4.5 & 1.2 & 1.6 & 8.9 & 3.5 & 3.3 & 9.0 & 7.6 & 9.4 & 14.3 \\
\bottomrule
\end{tabular}
}
\caption{Probability (\%) that decomposed sub-queries in retrieval fail to retrieve any positive document in the top 50.}
\label{tab:subq-recall-failure-transposed}
\end{table*}

\begin{table*}[h]
\centering
\small
\setlength{\tabcolsep}{8pt}
\renewcommand{\arraystretch}{1.15}
\resizebox{\textwidth}{!}{
\begin{tabular}{lcccccccccc}
\toprule
\textbf{Dataset} & \textbf{Q1} & \textbf{Q2} & \textbf{Q3} & \textbf{Q4} & \textbf{Q5} & \textbf{Q6} & \textbf{Q7} & \textbf{Q8} & \textbf{Q9} & \textbf{Q10} \\
\midrule
\textbf{Books} & 7.9 & 4.2 & 2.5 & 2.5 & 1.1 & 1.3 & 0.6 & 1.1 & 1.5 & 1.1 \\
\textbf{People} & 3.6 & 0.2 & 0.2 & 0.0 & 0.0 & 0.0 & 0.0 & 0.0 & 0.0 & 0.0 \\
\textbf{Movies} & 23.0 & 8.8 & 2.2 & 0.8 & 0.2 & 0.2 & 0.0 & 0.2 & 0.0 & 0.0 \\
\textbf{Medical Cases} & 12.5 & 3.3 & 0.4 & 0.2 & 0.0 & 0.0 & 0.0 & 0.0 & 0.0 & 0.0 \\
\textbf{Legal Documents} & 4.5 & 1.2 & 1.6 & 0.0 & 0.0 & 0.3 & 0.0 & 0.3 & 0.0 & 0.0 \\
\bottomrule
\end{tabular}
}
\caption{Probability (\%) that baseline retrieval fails to retrieve any positive document in the top 50.}
\label{tab:baseline-recall-failure-transposed}
\end{table*}

\subsection{Detailed Win Rate In Retrieval}
Figure~\ref{fig:recall_failure} reports the average win rate in retrieval process across the five datasets. 
For completeness, we provide the detailed results for each dataset (Books, Movies, People, Medical Cases, and Legal Documents) in the Appendix.

Tables ~\ref{tab:win_rate_comparison} present the detailed results for each dataset in MultiConIR.

\begin{table*}[h]
\centering
\small
\setlength{\tabcolsep}{8pt}
\renewcommand{\arraystretch}{1.15}
\resizebox{\textwidth}{!}{
\begin{tabular}{llcccccccccc}
\toprule
\textbf{Dataset} & \textbf{Setting} & \textbf{Q1} & \textbf{Q2} & \textbf{Q3} & \textbf{Q4} & \textbf{Q5} & \textbf{Q6} & \textbf{Q7} & \textbf{Q8} & \textbf{Q9} & \textbf{Q10} \\
\midrule
\multirow{2}{*}{\textbf{Books}}
& Baseline   & 74.7 & 68.3 & 64.9 & 59.2 & 57.3 & 56.4 & 55.2 & 51.2 & 53.2 & 51.1 \\
& Decomp     & 74.7 & 68.3 & 64.9 & 57.5 & 56.7 & 53.2 & 54.8 & 51.6 & 53.0 & 48.1 \\
\midrule
\multirow{2}{*}{\textbf{Movies}}
& Baseline   & 84.0 & 84.6 & 79.0 & 74.5 & 72.2 & 69.3 & 67.6 & 67.5 & 63.6 & 68.1 \\
& Decomp     & 84.0 & 84.6 & 79.0 & 68.5 & 69.1 & 65.4 & 63.3 & 59.4 & 65.3 & 64.2 \\
\midrule
\multirow{2}{*}{\textbf{People}}
& Baseline   & 74.1 & 71.4 & 69.3 & 68.3 & 65.4 & 67.0 & 63.3 & 64.6 & 59.0 & 66.3 \\
& Decomp     & 74.1 & 71.4 & 69.3 & 66.8 & 65.4 & 66.5 & 62.0 & 60.4 & 59.8 & 58.8 \\
\midrule
\multirow{2}{*}{\textbf{Medical Cases}}
& Baseline   & 84.6 & 77.2 & 74.3 & 70.2 & 65.2 & 67.3 & 67.1 & 66.5 & 66.4 & 56.8 \\
& Decomp     & 84.6 & 77.2 & 74.3 & 70.9 & 65.6 & 66.6 & 61.6 & 63.7 & 62.3 & 61.5 \\
\midrule
\multirow{2}{*}{\textbf{Legal Documents}}
& Baseline   & 86.2 & 75.1 & 73.9 & 67.9 & 63.2 & 62.8 & 63.4 & 58.0 & 53.0 & 49.0 \\
& Decomp     & 86.2 & 75.1 & 73.9 & 64.0 & 57.3 & 63.0 & 56.7 & 51.0 & 53.3 & 51.0 \\
\bottomrule
\end{tabular}
}
\caption{Win Rate (\%) between Baseline and Decomposition in retrieval methods across different datasets.}
\label{tab:win_rate_comparison}
\end{table*}

\clearpage %

\subsection{Detailed Win Rate In Reranking}
Figure~\ref{fig:win_rate_positive_document_reranking} reports the average win rate in reranking process across the five datasets. 
For completeness, we provide the detailed results for each dataset (Books, Movies, People, Medical Cases, and Legal Documents) in the Appendix.

Tables ~\ref{tab:winrate-appendix-reranking-qwen8b-results} present the detailed results for each dataset in MultiConIR.

\begin{table*}[h]
\centering
\small
\setlength{\tabcolsep}{8pt}
\renewcommand{\arraystretch}{1.15}
\resizebox{\textwidth}{!}{
\begin{tabular}{llcccccccccc}
\toprule
Dataset & Setting & Q1 & Q2 & Q3 & Q4 & Q5 & Q6 & Q7 & Q8 & Q9 & Q10 \\
\midrule
\multicolumn{12}{c}{\textbf{Win Rate}} \\
\midrule
\multirow{4}{*}{\textbf{Books}}
& Orig. Base    & 73.6 & 68.3 & 65.9 & 58.8 & 57.1 & 55.8 & 55.1 & 51.3 & 52.8 & 51.4 \\
& Pure Rerank   & 93.2 & 93.9 & 88.8 & 86.2 & 84.5 & 83.9 & 82.2 & 84.5 & 87.3 & 89.7 \\
& Decomp-Sum    & 93.2 & 93.9 & 88.8 & 88.0 & 86.0 & 85.0 & 83.7 & 88.2 & 90.4 & 90.5 \\
& Decomp-RRF    & 93.2 & 93.9 & 88.8 & 87.1 & 86.2 & 85.0 & 81.1 & 87.8 & 88.4 & 91.0 \\
\multirow{4}{*}{\textbf{Movies}}
& Orig. Base    & 82.0 & 83.9 & 78.6 & 74.7 & 72.3 & 69.3 & 67.7 & 67.4 & 63.6 & 68.1 \\
& Pure Rerank   & 89.9 & 90.6 & 88.7 & 89.2 & 88.1 & 85.8 & 85.0 & 81.3 & 79.8 & 76.0 \\
& Decomp-Sum    & 89.9 & 90.6 & 88.7 & 87.1 & 84.6 & 85.6 & 83.1 & 86.2 & 89.2 & 78.5 \\
& Decomp-RRF    & 89.9 & 90.6 & 88.7 & 84.0 & 83.0 & 84.0 & 79.7 & 81.9 & 82.5 & 74.6 \\
\multirow{4}{*}{\textbf{People}}
& Orig. Base    & 74.4 & 71.1 & 69.2 & 68.5 & 65.4 & 67.0 & 63.3 & 64.6 & 59.0 & 66.3 \\
& Pure Rerank   & 92.4 & 93.9 & 87.7 & 88.5 & 85.2 & 81.3 & 76.2 & 71.1 & 69.7 & 70.6 \\
& Decomp-Sum    & 92.4 & 93.9 & 87.7 & 91.0 & 85.4 & 89.4 & 82.4 & 85.5 & 88.0 & 89.5 \\
& Decomp-RRF    & 92.4 & 93.9 & 87.7 & 90.5 & 88.6 & 89.4 & 87.6 & 84.0 & 86.0 & 87.4 \\
\multirow{4}{*}{\textbf{Medical Cases}}
& Orig. Base    & 84.0 & 77.5 & 74.3 & 70.2 & 65.2 & 67.3 & 67.1 & 66.5 & 66.4 & 56.8 \\
& Pure Rerank   & 87.6 & 92.3 & 91.3 & 88.8 & 82.3 & 79.0 & 75.9 & 68.9 & 69.5 & 68.5 \\
& Decomp-Sum    & 87.6 & 92.3 & 91.3 & 92.5 & 89.6 & 90.5 & 85.3 & 88.4 & 90.2 & 88.3 \\
& Decomp-RRF    & 87.6 & 92.3 & 91.3 & 92.8 & 89.1 & 89.4 & 84.6 & 86.3 & 86.6 & 87.1 \\
\multirow{4}{*}{\textbf{Legal Documents}}
& Orig. Base    & 84.6 & 75.0 & 73.6 & 67.7 & 63.3 & 62.7 & 63.7 & 59.4 & 56.3 & 55.1 \\
& Pure Rerank   & 88.4 & 83.4 & 81.4 & 80.6 & 76.0 & 74.6 & 74.4 & 70.3 & 67.8 & 68.9 \\
& Decomp-Sum    & 88.4 & 83.4 & 81.4 & 83.3 & 81.4 & 81.7 & 82.1 & 82.5 & 82.0 & 85.4 \\
& Decomp-RRF    & 88.4 & 83.4 & 81.4 & 81.4 & 78.5 & 77.8 & 77.6 & 78.6 & 79.8 & 81.3 \\
\bottomrule
\end{tabular}
}
\caption{Detailed win rate (\%) results after reranking for each dataset.}
\label{tab:winrate-appendix-reranking-qwen8b-results}
\end{table*}

\newpage

\subsection{Detailed Positive Document Rank In Reranking}
Figure~\ref{fig:win_rate_positive_document_reranking} reports the average positive document rank in reranking process across the five datasets. 
For completeness, we provide the detailed results for each dataset (Books, Movies, People, Medical Cases, and Legal Documents) in the Appendix.

Tables ~\ref{tab:posposition-appendix-reranking} present the detailed results for each dataset in MultiConIR.

\begin{table*}[h]
\centering
\small
\setlength{\tabcolsep}{4pt}
\renewcommand{\arraystretch}{1.15}
\resizebox{\textwidth}{!}{
\begin{tabular}{llcccccccccc}
\toprule
Dataset & Setting & Q1 & Q2 & Q3 & Q4 & Q5 & Q6 & Q7 & Q8 & Q9 & Q10 \\
\midrule
\multicolumn{12}{c}{\textbf{Positive Document Average Position}} \\
\midrule
\multirow{4}{*}{\textbf{Books}}
& Orig. Base    & 8.46 & 8.19 & 7.48 & 7.03 & 6.70 & 6.62 & 6.24 & 6.37 & 6.20 & 6.58 \\
& Pure Rerank   & 5.72 & 5.31 & 5.20 & 5.13 & 5.12 & 5.13 & 5.06 & 5.06 & 4.67 & 3.10 \\
& Decomp-Sum    & 5.72 & 5.31 & 5.20 & 4.89 & 4.91 & 4.75 & 4.60 & 4.54 & 4.15 & 2.50 \\
& Decomp-RRF    & 5.72 & 5.31 & 5.20 & 5.22 & 5.14 & 5.08 & 5.13 & 5.06 & 4.68 & 2.48 \\
\multirow{4}{*}{\textbf{Movies}}
& Orig. Base    & 11.36 & 8.02 & 5.96 & 5.29 & 4.85 & 4.76 & 4.66 & 4.48 & 4.49 & 4.28 \\
& Pure Rerank   & 6.13 & 5.19 & 4.88 & 4.57 & 4.37 & 4.15 & 3.95 & 3.90 & 3.93 & 3.98 \\
& Decomp-Sum    & 6.13 & 5.19 & 4.88 & 4.57 & 4.39 & 4.03 & 3.78 & 3.51 & 3.05 & 2.61 \\
& Decomp-RRF    & 6.13 & 5.19 & 4.88 & 5.05 & 4.78 & 4.52 & 4.49 & 4.22 & 3.82 & 3.18 \\
\multirow{4}{*}{\textbf{People}}
& Orig. Base    & 8.01 & 6.62 & 6.07 & 5.49 & 5.37 & 5.29 & 5.23 & 5.13 & 5.11 & 4.81 \\
& Pure Rerank   & 5.51 & 5.26 & 5.12 & 4.95 & 4.82 & 4.75 & 4.75 & 4.70 & 4.80 & 4.64 \\
& Decomp-Sum    & 5.51 & 5.26 & 5.12 & 4.85 & 4.78 & 4.49 & 4.31 & 4.13 & 3.66 & 2.22 \\
& Decomp-RRF    & 5.51 & 5.26 & 5.12 & 5.14 & 4.96 & 4.77 & 4.70 & 4.52 & 4.08 & 2.31 \\
\multirow{4}{*}{\textbf{Medical Cases}}
& Orig. Base    & 11.23 & 8.00 & 6.71 & 5.66 & 5.39 & 5.13 & 4.96 & 4.78 & 4.42 & 4.68 \\
& Pure Rerank   & 6.58 & 5.36 & 4.96 & 4.78 & 4.74 & 4.65 & 4.69 & 4.83 & 4.81 & 5.01 \\
& Decomp-Sum    & 6.58 & 5.36 & 4.96 & 4.60 & 4.41 & 4.07 & 3.84 & 3.53 & 3.19 & 2.29 \\
& Decomp-RRF    & 6.58 & 5.36 & 4.96 & 4.99 & 4.78 & 4.52 & 4.57 & 4.30 & 4.08 & 2.52 \\
\multirow{4}{*}{\textbf{Legal Documents}}
& Orig. Base    & 9.05 & 7.02 & 6.17 & 5.69 & 5.38 & 5.19 & 4.95 & 5.03 & 4.97 & 5.12 \\
& Pure Rerank   & 5.99 & 5.64 & 5.31 & 5.17 & 5.13 & 5.06 & 4.98 & 5.01 & 5.01 & 4.98 \\
& Decomp-Sum    & 5.99 & 5.64 & 5.31 & 4.99 & 4.94 & 4.76 & 4.37 & 4.19 & 4.04 & 2.90 \\
& Decomp-RRF    & 5.99 & 5.64 & 5.31 & 5.34 & 5.23 & 5.05 & 4.89 & 4.80 & 4.50 & 3.23 \\
\bottomrule
\end{tabular}
}
\caption{Detailed positive document position results after reranking for each dataset.}
\label{tab:posposition-appendix-reranking}
\end{table*}

\clearpage
\raggedbottom 
\onecolumn

\section{Additional Details on Reranking Setup}
\label{app:logit_reranking}

\subsection{Why We Use Pointwise Reranking Instead of Listwise Reranking}

Pointwise and listwise reranking are two standard paradigms for reranking. While listwise reranking has shown strong performance in many ranking benchmarks, applying it to large candidate pools is computationally expensive due to the long-context interaction among multiple documents. In addition, long-context listwise reranking is more susceptible to positional biases such as the ``lost in the middle'' effect~\cite{liu2024lost}, where documents located in the middle of the context receive less effective attention.

In our setting, query decomposition introduces multiple sub-queries for each original query. Combining listwise reranking with decomposition would substantially increase context length and computational overhead, since all candidate documents and multiple sub-queries would need to be jointly processed. To avoid these issues, we adopt a pointwise reranking strategy, where each candidate document is independently scored against each decomposed sub-query.

\subsection{Preliminary Token-Based Reranking Exploration}

In our preliminary experiments, we explored a zero-shot \textbf{token-based reranking} approach using Qwen3-0.6B. In this setup, the model was prompted to directly generate a discrete numerical relevance score (e.g., 0--10) for each query-document pair.

However, we found that this generative scoring strategy was unstable and consistently underperformed the original retrieval baseline. We attribute this degradation to two main factors:

\begin{itemize}
    \item \textbf{Generation instability:} autoregressive generation introduces decoding variance and formatting inconsistency for numerical outputs.
    
    \item \textbf{Weak score calibration:} discrete generated scores are not well calibrated across different query-document pairs, making score aggregation less reliable.
\end{itemize}

To address these limitations, we transitioned to a \textbf{logit-based reranking} formulation using the Qwen3-Reranker series (0.6B, 4B, and 8B). Specifically, we formulate reranking as a binary classification task. Instead of relying on generated text outputs, we directly extract the logits corresponding to the ``Yes'' and ``No'' tokens from the final prediction layer. These logits are normalized using a softmax function to produce a continuous relevance score between 0 and 1:

\begin{equation}
P(\text{Yes}|q,d)=
\frac{
\exp(z_{\text{Yes}})
}{
\exp(z_{\text{Yes}})+\exp(z_{\text{No}})
}
\end{equation}

where $z_{\text{Yes}}$ and $z_{\text{No}}$ denote the logits of the corresponding tokens.

Table~\ref{tab:scoring-comparison-books} compares the performance of the token-based and logit-based scoring strategies on the Books dataset. The results show that the logit-based formulation consistently outperforms both the token-based approach and the original retrieval baseline.

\begin{table}[H]
\centering
\small
\setlength{\tabcolsep}{8pt}
\renewcommand{\arraystretch}{1.2}
\begin{tabular}{lccc}
\toprule
\textbf{Query} & \textbf{Baseline} & \textbf{Token-Based} & \textbf{Logit-Based} \\
\midrule
Query 1  & 0.753 & 0.696 & 0.852 \\
Query 2  & 0.761 & 0.722 & 0.880 \\
Query 3  & 0.778 & 0.752 & 0.867 \\
Query 4  & 0.758 & 0.750 & 0.821 \\
Query 5  & 0.745 & 0.735 & 0.798 \\
Query 6  & 0.699 & 0.682 & 0.737 \\
Query 7  & 0.661 & 0.645 & 0.680 \\
Query 8  & 0.581 & 0.564 & 0.596 \\
Query 9  & 0.494 & 0.463 & 0.514 \\
Query 10 & 0.367 & 0.345 & 0.438 \\
\bottomrule
\end{tabular}
\caption{Performance comparison (NDCG@10) of different scoring mechanisms on the Books dataset. The logit-based approach consistently resolves the degradation caused by token-based scoring and significantly surpasses the initial retrieval baseline.}
\label{tab:scoring-comparison-books}
\end{table}
\subsection{Implementation Details of Logit-Based Reranking}

In our logit-based reranking phase (using the Qwen3-Reranker series), the model evaluates the relevance of a document to a given query using a strictly structured instructional prompt. The exact input is concatenated as follows:

\begin{verbatim}
<Instruct>: {instruction}
<Query>: {query}
<Document>: {doc}
\end{verbatim}

Instead of decoding the full generated sequence, we directly extract the logits corresponding to the positive and negative label tokens at the final prediction layer. In our implementation, these are represented by $\texttt{token\_true\_id}$ (for ``Yes'') and $\texttt{token\_false\_id}$ (for ``No''). The final relevance score is computed by applying a log-softmax formulation over these two specific logits:

\begin{equation}
\text{score} =
\frac{\exp(L_{\text{true}})}
{\exp(L_{\text{true}}) + \exp(L_{\text{false}})}
\end{equation}

where $L$ represents the raw logit value for the respective token. This methodology ensures the model leverages its internal probability distribution, providing a well-calibrated, continuous probability score that serves as a highly deterministic signal for our End-to-End Pipeline.

\newpage

\section{Semantic Dilution in Decomposed Sub-queries}
\label{app:recall_failure}
We present an example from the Books dataset to illustrate how decomposition weakens semantic constraints and leads to easy negative retrieval. The original query specifies 8 conditions and is partitioned into sub-queries. We examine one such sub-query, which retains 3 constraints. Under this sub-query, the dense retriever assigns higher similarity score to an easy negative (0.650) than to the true positive (0.576).

\begin{tcolorbox}[
    colback=gray!5,
    colframe=gray!50,
    title=\textbf{Case Study: Semantic Dilution Example},
    fonttitle=\bfseries,
    boxrule=0.5pt,
    arc=2pt,
]
\textbf{\# Original Query (8 conditions)} \\
Find a book that fits: 1. Authored by Steve Chandler, 2. Helps with personal change, 3. Contains humor and reassurance, 4. Recommended by peers, 5. Offers an action plan, 6. Includes feedback-based methods, 7. Uses coaching techniques, 8. Banishes negative thoughts.

\vspace{0.5em}
\textbf{\# Decomposed Sub-query (3 conditions retained)} \\
Find a book that satisfies these conditions: Contains humor and reassurance. Recommended by peers. Offers an action plan.

\vspace{0.5em}
\textbf{\# Positive Document (similarity: 0.576)} \\
``\,`If you take the best of Wayne Dyer and add it to the best of Anthony Robbins, what you would have would only be half as good as Steve Chandler.'---Dale Dauten, Chicago Tribune `Some books that can help you awaken and begin to change are ones by Steve Chandler, who, I am reading lately. Great stuff. I'm becoming a fan of Steve Chandler.'---Joe Vitale, best-selling author of The Attraction Factor and contributor to The Secret `Steve Chandler lights you up with the glow of his internal neon. \textbf{[He is] one of the funniest men I've ever known---what he proposes is so rock solid and reassuring.}'---Lisa Schnebly, The Arizona Republic `100 Ways to Motivate Yourself is wonderful, inspirational, honest, and courageous. It speaks from every page. \textbf{It is definitely a book I will recommend to my clients and friends.}'---Devers Branden, coauthor of What Love Asks of Us With the third refreshed edition of 100 Ways to Motivate Yourself, \textbf{Steve Chandler helps you create an action plan for living your vision}, in business and in life. It features 100 proven methods to positively change the way you think and act---methods based on feedback from the hundreds of thousands of corporate and public seminar attendees Chandler speaks to each year. The book now also includes techniques and breakthroughs he has created for individual coaching clients. 100 Ways to Motivate Yourself will help you break through the negative barriers and banish the pessimistic thoughts that are preventing you from fulfilling your lifelong goals and dreams. This edition also contains new mental and spiritual techniques that give readers more immediate access to action and results in their lives. Steve Chandler invites you to maintain your current mindset and stick to conventional methods, leading to minimal transformation and delaying your progress toward your objectives.''

\vspace{0.5em}
\textbf{\# Easy Negative Document (similarity: 0.650)} \\
``You can wait for the right moment before setting out on your dreams. You can learn to successfully deal with any difficult times or negative people you may encounter. You can unlock enthusiasm, determination, and energy using the keys of an optimistic point of view: releasing the past, keeping the future in mind, and taking everything one day at a time. Day by day, this book addresses a different topic relating to issues you may face as you work to improve your mental outlook. Beginning with a `game plan' and mapping out where you want to go, you'll gain insight about working through obstacles and measuring your progress. You'll also be reminded about the value of appreciating your present blessings and looking forward to a hopeful future. You'll find practical advice as well as writings specifically chosen to inspire and motivate you on your journey. The empowering message of this book is also a simple one: You have the power in your hands today to be happier, more creative, and more at peace with yourself and everyone around you. Take this week out of your life and begin to build a positive attitude that will reward you in wonderful ways and make a lasting difference in this world. Category: Self-help, Personal Growth, General''
\end{tcolorbox}

\section{Prompt Template for GPT-5.3 in SSRB Generalization Analysis}
\label{app:prompt}

In this appendix, we detail the exact prompt instructions provided to GPT-5.3 to perform fixed-size query decomposition on the SSRB dataset. The prompt is designed to strictly enforce structural boundaries while preserving the semantic integrity of the original multi-condition queries.

\vspace{0.5cm}

\begin{tcolorbox}[colback=gray!5, colframe=black!70, title=\textbf{System Prompt for Query Decomposition}, arc=2mm, boxrule=0.5pt, left=3mm, right=3mm, top=2mm, bottom=2mm]
\small

\textbf{\# Role \& Task} \\
You are an expert in complex, multi-condition information retrieval. Your task is to decompose a given complex search query into distinct, independent sub-queries to optimize downstream retrieval performance.

\vspace{2mm}
\textbf{\# Strict Rules for Decomposition}
\begin{enumerate}
    \item \textbf{Decomposition Limit:} Split the original query into sub-queries. If the original query only contains 2-3 conditions, output the original query as a single sub-query without splitting.
    \item \textbf{Condition Threshold:} Each generated sub-query MUST contain 2-3 distinct search conditions or constraints.
    \item \textbf{No Information Overlap:} The semantic conditions must be mutually exclusive across sub-queries. Do not duplicate specific constraints or information across the decomposed segments.
    \item \textbf{Preserve Original Wording:} Extract and segment the text while strictly preserving the original phrasing, vocabulary, and sentence structure as much as possible. Avoid aggressively paraphrasing, rewriting, or hallucinating new information. Ensure each segment remains a coherent sentence.
\end{enumerate}

\vspace{2mm}
\textbf{\# Output Format} \\
You must strictly output the decomposed sub-queries in the following format without any additional conversational text:\\
\texttt{[Original\_ID]\_subq\_[Index], "[Sub-query text]"}

\vspace{2mm}
\textbf{\# Example} \\
\textbf{Input:} \\
\texttt{product\_search--camera\_search--q--1347, "I'm planning a hiking trip and need a durable, high-quality camera that can handle outdoor conditions. I'm looking for a camera with a resolution of at least 20 megapixels and one that's known for its excellent image quality, something with a really good reputation amongst photography enthusiasts."}

\vspace{2mm}
\textbf{Output:} \\
\texttt{product\_search--camera\_search--q--1347\_subq\_1, "I'm planning a hiking trip and need a durable, high-quality camera that can handle outdoor conditions. I'm looking for a camera with a resolution of at least 20 megapixels"} \\
\texttt{product\_search--camera\_search--q--1347\_subq\_2, "I'm looking for a high quality camera and one that's known for its excellent image quality, something with a really good reputation amongst photography enthusiasts."}
\end{tcolorbox}

\newpage

\section{Sensitivity to the RRF Hyperparameter}
\label{app:rrf_sensitivity}

RRF introduces a smoothing constant $k$ that controls the contribution of rank positions during score aggregation. To examine whether RRF is sensitive to this hyperparameter, we vary $k$ from 10 to 120 and report the resulting performance in Table~\ref{tab:rrf_sensitivity}.

\begin{table}[h]
\centering
\setlength{\tabcolsep}{6pt}
\renewcommand{\arraystretch}{1.05}
\begin{tabular}{cccc}
\toprule
\textbf{$k$} &
\textbf{NDCG@10} &
\textbf{Recall@10} &
\textbf{Recall@50} \\
\midrule
10  & 0.5935 & 0.7706 & 0.9407 \\
20  & 0.5995 & 0.7753 & 0.9396 \\
40  & 0.6007 & 0.7757 & 0.9390 \\
60  & 0.5994 & 0.7737 & 0.9387 \\
80  & 0.5983 & 0.7718 & 0.9376 \\
100 & 0.5974 & 0.7703 & 0.9361 \\
120 & 0.5968 & 0.7692 & 0.9348 \\
\bottomrule
\end{tabular}
\caption{Sensitivity analysis of the RRF smoothing parameter $k$.}
\label{tab:rrf_sensitivity}
\end{table}

As shown in Table~\ref{tab:rrf_sensitivity}, performance remains relatively stable across a broad range of $k$ values. NDCG@10 varies only from 0.5935 to 0.6007, while Recall@10 and Recall@50 also show only minor fluctuations. These results indicate that RRF is not highly sensitive to the choice of $k$ in our setting.

\end{document}